\documentclass[12pt]{article}
\usepackage{graphicx}
\def\hybrid{\topmargin 0pt      \oddsidemargin 0pt
        \headheight 0pt \headsep 0pt
        \voffset=-0.5cm
        \textwidth 6.25in       
        \textheight 9.5in       
        \marginparwidth 0.0in
        \parskip 5pt plus 1pt   \jot = 1.5ex}
\catcode`\@=11
\def\marginnote#1{}

\newcount\hour
\newcount\minute
\newtoks\amorpm
\hour=\time\divide\hour by60
\minute=\time{\multiply\hour by60 \global\advance\minute by-\hour}
\edef\standardtime{{\ifnum\hour<12 \global\amorpm={am}%
        \else\global\amorpm={pm}\advance\hour by-12 \fi
        \ifnum\hour=0 \hour=12 \fi
        \number\hour:\ifnum\minute<10 0\fi\number\minute\the\amorpm}}
\edef\militarytime{\number\hour:\ifnum\minute<10 0\fi\number\minute}

\def\draftlabel#1{{\@bsphack\if@filesw {\let\thepage\relax
   \xdef\@gtempa{\write\@auxout{\string
      \newlabel{#1}{{\@currentlabel}{\thepage}}}}}\@gtempa
   \if@nobreak \ifvmode\nobreak\fi\fi\fi\@esphack}
        \gdef\@eqnlabel{#1}}
\def\@eqnlabel{}
\def\@vacuum{}
\def\draftmarginnote#1{\marginpar{\raggedright\scriptsize\tt#1}}
\def\draftlabel#1{{\@bsphack\if@filesw {\let\thepage\relax
   \xdef\@gtempa{\write\@auxout{\string
      \newlabel{#1}{{\@currentlabel}{\thepage}}}}}\@gtempa
   \if@nobreak \ifvmode\nobreak\fi\fi\fi\@esphack}
        \gdef\@eqnlabel{#1}}
\def\@eqnlabel{}
\def\@vacuum{}
\def\draftmarginnote#1{\marginpar{\raggedright\scriptsize\tt#1}}

\def\draft{\oddsidemargin -.5truein
        \def\@oddfoot{\sl preliminary draft \hfil
        \rm\thepage\hfil\sl\today\quad\militarytime}
        \let\@evenfoot\@oddfoot \overfullrule 3pt
        \let\label=\draftlabel
        \let\marginnote=\draftmarginnote
   \def\@eqnnum{(\theequation)\rlap{\kern\marginparsep\tt\@eqnlabel}%
\global\let\@eqnlabel\@vacuum}  }

\def\numberbysection{\@addtoreset{equation}{section}
        \def\theequation{\thesection.\arabic{equation}}}

\def\underline#1{\relax\ifmmode\@@underline#1\else
        $\@@underline{\hbox{#1}}$\relax\fi}

\def\titlepage{\@restonecolfalse\if@twocolumn\@restonecoltrue\onecolumn
     \else \newpage \fi \thispagestyle{empty}\c@page\z@
        \def\thefootnote{\fnsymbol{footnote}} }

\def\endtitlepage{\if@restonecol\twocolumn \else  \fi
        \def\thefootnote{\arabic{footnote}}
        \setcounter{footnote}{0}}  
\relax

\numberbysection
\hybrid

\newfont{\Bbb}{msbm10 scaled 1\@ptsize00}
\newcommand{\CC}{\mbox{\Bbb C}}

\newfont{\Bbbb}{msbm7 scaled 1\@ptsize00}
\newcommand{\cc}{\raise-1pt\hbox{$\mbox{\Bbbb C}$}}
\newcommand{\zz}{\raise-1pt\hbox{$\mbox{\Bbbb Z}$}}

\def\beq{\begin{equation}}
\def\eeq{\end{equation}}
\def\p{\partial}

\def\DD{{\sf D}}
\def\Dc{\CC \! \setminus \! {\sf D}}

\def\lbracket{\left <}
\def\rbracket{\right >}

\def\pvint{\int_{\Gamma} \hspace{-13pt} {\bf -}}

\begin{document}

\begin{titlepage}

\title{Dyson gas on a curved contour}

\author{P. Wiegmann\thanks{Kadanoff Center for Theoretical Physics, University
of Chicago, 5640 South Ellis Ave, Chicago, IL 60637, USA; 
}
\and
A.~Zabrodin\thanks{
Skolkovo Institute of Science and Technology, 143026, Moscow, Russia and
National Research University Higher School of Economics,
20 Myasnitskaya Ulitsa,
Moscow 101000, Russia and
ITEP NRC KI, 25
B.Cheremushkinskaya, Moscow 117218, Russia;
e-mail: zabrodin@itep.ru}}

\date{November 2021}
\maketitle

\vspace{-7cm} \centerline{ \hfill ITEP-TH-30/21}\vspace{7cm}

\begin{abstract}

We introduce and study a model of a logarithmic gas with inverse temperature
$\beta$ on an arbitrary smooth closed contour in the plane. This model
generalizes Dyson's gas (the $\beta$-ensemble) on the unit circle.
We compute the non-vanishing terms of the large $N$ expansion of the free energy
($N$ is the number of particles) by iterating the ``loop equation'' 
that is the Ward identity with respect to reparametrizations and dilatation 
of the contour.
We show that the main contribution to the free energy is expressed through
the spectral determinant of the Neumann jump operator associated with the contour,
or equivalently through the Fredholm determinant of the Neumann-Poincare
(double layer) operator. This result connects the statistical mechanics of 
the Dyson gas to the
spectral geometry of the interior and exterior domains of the supporting contour.

\end{abstract}

\vfill

\end{titlepage}

\tableofcontents

\section{Introduction}

The logarithmic gases were introduced by F.Dyson in the seminal papers 
\cite{Dyson}, where it was shown that eigenvalues of random matrices can be represented
as a statistical ensemble of charged particles with 2D Coulomb (logarithmic) interaction
at inverse temperature $\beta >0$. Particles can be 
located either on a circle or on an interval of a straight line.
The partition function of the circular Dyson's ensemble is
\beq\label{int1}
Z_N = \oint_{|z_1|=1}\ldots \oint_{|z_N|=1}|\Delta_N (\{z_i\})|^{2\beta}|dz_1|\ldots |dz_N|,
\quad z_j =e^{i\theta_j}, \quad |dz_j|=d\theta_j.
\eeq
Here
\beq\label{int2}
\Delta_N (\{z_i\})=\prod_{j<k}^N (z_j-z_k)
\eeq
is the Vandermonde determinant. 
A comprehensive description of logarithmic gases
can be found in \cite{Forrester}.

The values $\beta =\frac{1}{2}, \, 1, \, 2$ are special.
In these cases, the logarithmic gases (on various supports) can be represented 
either through free fermions or through some random matrix ensembles.
Furthermore, for these special values of $\beta$ the 
log-gases possess an integrable structure. 
In particular, the $\beta=1$ partition function is the tau-function of the 2D 
Toda lattice hierarchy (see (\ref{ZNab}), (\ref{ZNc}) below).
Positive integer values of $\beta$ are also special. In particular,
the two-point correlation function then permits a $\beta$-dimensional
integral representation valid for any $N$. 
Other values of $\beta$ do not seem to possess special properties. 
Nevertheless, a few  particular integrals where computed explicitly for any values 
of $\beta$. In particular, the integral (\ref{int1}) equals
$$ 
Z_N=(2\pi )^N \frac{\Gamma (1+N\beta )}{\Gamma ^N (1+\beta )}
$$ 
(see \cite{Dyson}).  
This is not the case for correlation functions, though.

In this paper, we introduce a natural generalization of the Dyson ensemble (\ref{int1}):
the logarithmic gas on an arbitrary closed contour
(a deformed circle) and study how the partition function and various 
correlation functions depend on the geometry of the contour.   
Namely, we consider the $\beta$-ensemble
of $N$ particles on a simple closed contour
$\Gamma$ on a plane. An early reference on this problem is \cite{Jancovici}
(section 7). 
For further convenience, we place the gas into $N$-independent 
external potential $W$. 
The partition function reads
\beq\label{ZN}
Z_N =\oint _{\Gamma} \! \ldots \! \oint _{\Gamma}
\prod_{i<j}|z_i -z_j |^{2\beta}
\prod_{k=1}^{N}
e^{W(z_k )}|dz_k |.
\eeq
Clearly, the dimension of $Z_N$ is
$[\mbox{length}]^{\beta N^2 +(1-\beta )N}$.
The real-valued function $W$ is defined
in a strip-like neighborhood of the contour and depends on both $z$ and $\bar z$
(we write $W=W(z)$ just for simplicity of the notation).
The meaning of the potential $W$ may be understood as a metric on the 
contour, $ds=e^W |dz|$. As a functional of $W$, the partition function is the generating 
functional of correlation functions of densities.
A similar problem having support on the complex plane, the 2D Dyson gas, 
had been considered by us earlier (see \cite{WZ03,WZ03a,WZ06,Z04}).

We are mostly interested in the large $N$ limit, $N\to \infty$, and in the 
large $N$ expansion of the free energy. Let us introduce the small parameter
\beq\label{hbar}
\hbar = \frac{1}{N},
\eeq
then in the large $N$ limit the behavior
of the partition function is of the form
$$Z_N = N!\, N^{(\beta -1)N} e^F,$$ where
\beq\label{large1}
F=\frac{F_0}{\hbar^2}+\frac{F_{1}}{\hbar} +F_2 +O(\hbar ), \quad \hbar \to 0.
\eeq
The factors $e^{F_0 /\hbar^2}$ and $e^{F_{1} /\hbar}$ carry
the dimensions
$[\mbox{length}]^{\beta N^2}$ and
$[\mbox{length}]^{(1-\beta ) N}$ respectively.  
We are interested in the contributions that do not vanish
as $\hbar \to 0$, i.e., in the 
first three terms 
$F_0$, $F_1$ and $F_2$ of the large $N$ expansion of the free energy $F$. 

We obtain explicit expressions for $F_0$, $F_1$ and $F_2$ in terms of the conformal map
$w_{\rm ext}(z)$ of the exterior domain bounded by the  contour $\Gamma$ 
to the exterior of the unit
circle (normalized as $w_{\rm ext}(\infty )=\infty$, $w_{\rm ext}'(\infty )>0$)
and the Neumann jump operator $\hat {\cal N}$ (which is defined below, 
see (\ref{Neumann})). 
The leading contribution, $F_0$, equals the electrostatic energy of a charged 2D conductor
of the shape $\Gamma$ (neutralized by a charge at infinity).
The next-to-leading contribution, $F_1$, reflects the entropy and the excluded volume 
(the effect that the place occupied by a particle is no longer accessible for others).
The most interesting  quantity is $F_2$. It is shown to be closely related to the spectral
theory  of the interior and exterior domains. Setting $W=0$, we have:
\beq\label{int5}
\begin{array}{l}
F_0=\beta \log r,
\\ \\
\displaystyle{
F_1= -(\beta -1)\log \frac{er}{2\pi \beta} + \log \frac{2\pi}{\Gamma (\beta )},}
\\ \\
F_2=\displaystyle{\frac{(\beta -1)^2}{8\pi \beta}\oint_{\Gamma}\log \rho_0 \, \hat {\cal N}
\log \rho_0 ds -\frac{1}{2}\log \frac{P}{2\beta} +\frac{1}{2}\, \log {\det} ' \hat {\cal N}.}
\end{array}
\eeq
Here $r=\Bigl (w_{\rm ext}'(\infty )\Bigr )^{-1}$ is the external conformal radius of the 
domain ${\sf D}$ bounded by the contour
(we also call it the conformal radius of the exterior domain $\CC \setminus {\sf D}$), 
$\displaystyle{\rho_0(z)=\frac{1}{2\pi} |w_{\rm ext}'(z)|}$ is the mean density of particles
in the leading order,
$P$ is the perimeter of the contour $\Gamma$,
$ds =|dz|$ is the line element
along the contour and ${\det}' \hat {\cal N}$ is the regularized spectral determinant 
of the Neumann jump operator with the zero mode removed. 
The result for $F_2$ can be naturally divided into ``classical'' part having an electrostatic
nature (the first term) 
and the ``quantum'' part which origin is similar to the gravitational anomaly 
of the free bosonic field localized on the contour. 
The quantum part can be also represented as the logarithm of the Fredholm 
determinant associated to the domain surrounded by the contour $\Gamma$. It is explicitly
given by equation (\ref{F2q1}). 
We also calculate one- and two-point correlation functions
of densities of the particles in the large $N$ limit.

Our main technical tool is the so-called loop equation, which is an exact relation 
connecting one-and two-point correlation functions. 
It follows from  the invariance of the
partition function under re-parameterizations of the contour (admissible changes of integration 
variables). The term ``loop equation'' goes back to the terminology used in the theory
of random matrices but long before it is known as
Bogoliubov-Born-Green-Kirkwood-Yvon hierarchy
in statistical mechanics or as Ward identity w.r.t. dilatations
in quantum field theory. 
Although this relation is to some extent tautological, it becomes
really meaningful when combined with the large $N$ expansion.
We show that the general form of the loop equation for contours
$\Gamma$ of arbitrary shape is a condition for the  boundary value of a holomorphic quantity
$T(z)$ closely related to the 
stress-energy tensor of a Gaussian field in 2D often utilized in conformal field theory. 
Its mean value $\bigl < T(z)\bigr >$
is a holomorphic function both in the interior and exterior domains with a jump
across the contour $\Gamma$.
The boundary condition for the 
mean value is equivalent
to the loop equation. It states that the jump of the off-diagonal 
tangential-normal 
component
$\Bigl [ \left < T_{sn}(z)\right >\Bigr ]_{\Gamma}=
\left < T_{sn}^{(+)}(z)\right >-\left < T_{sn}^{(-)}(z)\right >$ 
across the contour ($z\in \Gamma$, $T^{(+)}$ is the value from inside 
and $T^{(-)}$ from outside) vanishes.  
Below we will use the notation $[\dots]_\Gamma$ for a jump across the contour.

Another technical tool is the rule of variation of the partition function under small
deformations of the contour. We show that this variation is expressed through the jump
$\Bigl [ \left < T_{nn}(z)\right >\Bigr ]_{\Gamma}$ of the normal-normal 
component of $\bigl <T(z)\bigr >$ across the contour:
\beq\label{int4}
\hbar^2 \delta \log Z_N =-\frac{1}{\pi \beta}
\oint_{\Gamma} \Bigl [ \left < T_{nn}(z)\right >\Bigr ]_{\Gamma} \delta n (z) \, |dz|.
\eeq
Here $\delta n(z)$ is the infinitesimal normal displacement of the contour at the point
$z\in \Gamma$. This formula, combined with the large $N$ expansion of the loop equation, 
allows one to find variations of $F_0$, $F_1$ and $F_2$ and then, having at hand the variations,
to restore them up to constant terms which do not depend on the shape of the contour. In principle, using the 
iterative procedure, one can also find the higher corrections to the free energy but 
the calculations become rather involved. This procedure is in parallel 
to the one which we used for the 2D Dyson gas in \cite{WZ06}. 

The organization of the paper is as follows. In Sec. 2 we introduce the distribution functions
and correlation functions of the main observables (density of charges, the potential 
created by them and an analog of the stress-energy tensor). 
In Sec. 3 we obtain general relations valid at a finite $N$ and
express the loop equation and the variation of the partition function in terms of
boundary values of $T(z)$. 
Sec. 4 is devoted to the 
large $N$ expansion. There we develop a systematic procedure 
of the large $N$ expansion and obtain 
explicit formulas for
the first three orders of the large $N$ expansion. 
In Sec. 5 we calculate correlation functions
of the analogs of vertex operators. There are also six Appendices.
In Appendix A we list a few simple but useful 
formulas for differential geometry of a contour. 
In Appendices B and C we review the Green functions
of Laplace operator and discuss the Neumann jump operator. 
In Appendix D we present some facts about the double layer potentials and Fredholm
eigenvalues associated with the contour $\Gamma$. 
Appendix E contains the technique of variation of contour integrals we used
in the  text. In Appendix F we discuss functional determinants of the Laplace operators
in interior and exterior domains.

\section{Some definitions and general relations}

The partition function of our model is defined by the integral 
(\ref{ZN}). We start with some  definitions and general relations. 

\subsection{Distribution functions}

Integrating in the r.h.s. of (\ref{ZN}) over a part
of the variables keeping
the others fixed, one can define
a set of quantities
which are proportional to the probability distributions to find
particles at given points.
For example,
\beq\label{R1}
R(a)=\frac{1}{Z_N}\,
e^{W(a)}
\oint _{\Gamma}|\Delta_{N-1}(z_i )|^{2\beta}
\prod_{j=1}^{N-1} |a -z_j |^{2\beta}
e^{W(z_j)}\, |d z_j |
\eeq
is the mean density of particles
while
\beq\label{R2}
\begin{array}{l}
\displaystyle{
R(a,b)=\frac{\hbar (N\! -\! 1)}{Z_N}\,\, |a\! -\! b|^{2\beta}\,
e^{W(a)+W(b)}} \\ \\
\displaystyle{ \phantom{aaaaaa}\times
\oint _{\Gamma} |\Delta_{N-2}(z_i )|^{2\beta}
\prod_{j=1}^{N-2} |a -z_j |^{2\beta}|b -z_j |^{2\beta}
e^{W(z_j)}\, |d z_j |}
\end{array}
\eeq
is the pair distribution function.
Obviously, $R(a,a)=0$ which means that two particles can not
be found in one and the same point.
The normalization is chosen so that the functions $R(a)$,
$R(a,b)$ are dimensionless and
$$
\oint _{\Gamma} R(z)\,  |d z| = 1 \,, \quad \quad
\oint _{\Gamma} R(a,z)\, |d z| =(N\! -\! 1)\, \hbar \, R(a).
$$

Let $A(z_1, \ldots , z_N)$ be an arbitrary symmetric
function of the variables $z_i$; its
mean value is defined as
\beq\label{mean}
\lbracket A\rbracket = \frac{1}{Z_N}
\oint _{\Gamma} \! \ldots \! \oint _{\Gamma} A(z_1, \ldots , z_N)
\prod_{i<j}|z_i -z_j |^{2\beta}
\prod_{k=1}^{N}
e^{W(z_k )}|dz_k |.
\eeq
Clearly, the mean values $\lbracket \sum_i f(z_i)\rbracket$
and $\lbracket \sum_{i\neq j} f(z_i , z_j)\rbracket$, etc are expressed
through the distribution functions as follows:
\beq\label{R3b}
\begin{array}{l}
\displaystyle{
\lbracket \sum_i f(z_i)\rbracket =
\frac{1}{\hbar}\oint _{\Gamma} R(z)f(z) |dz|,}
\\ \\
\displaystyle{
\lbracket \sum_{i\neq j} f(z_i, z_j)\rbracket =
\frac{1}{\hbar ^2}\oint _{\Gamma} \! \oint _{\Gamma}
R(z, \zeta )f(z, \zeta ) |dz| |d\zeta |\,.}
\end{array}
\eeq

\subsection{Density, potential and their correlation functions}

The most important observables in the model are density
of particles and
2D Coulomb field created by them.
The density is defined as
\beq\label{rho}
\rho (z)=\hbar \sum_{i=1}^{N} \delta_{\Gamma}(z,z_i ), 
\eeq
where $\delta_{\Gamma}(z,z')$ is the $\delta$-function
on $\Gamma$ defined by the property
$\oint_{\Gamma}f(z)\delta_{\Gamma}(z,a)|dz|=f(a)$
for $a\in \Gamma$ and a continuous function $f$.
The field
\beq\label{phi}
\varphi (z)=-\beta \hbar \sum_i \log |z-z_i|^2
=-2\beta \oint_{\Gamma}\log |z-\xi | \, \rho (\xi)|d\xi |
\eeq
is the 2D Coulomb potential created by the particles.
Equation (\ref{phi})
says that $\varphi (z)$ is the
potential of a simple layer with
density $\rho$. The simple layer potential $\varphi (z)$ is harmonic everywhere
in the plane except for the contour $\Gamma$,
where it is continuous across the boundary but has a jump
of the normal derivative 
\beq\label{jump}
[\p_n \varphi (z)]_{\Gamma}=
\p_n ^{+}\varphi (z)-\p_n ^{-}\varphi (z)=4\pi \beta \rho (z),
\quad z\in \Gamma ,
\eeq
where $\p_n ^{+}$ and $\p_n ^{-}$ are normal derivatives inside
and outside the contour $\Gamma$ respectively
(with the unit normal vector
looking outward in both cases). At infinity the field $\varphi$ 
behaves as
\beq\label{logsing}
\varphi (z)=-2\beta \log |z| + O(|z|^{-1}).
\eeq
Note the absence of the $O(1)$ term.

Equation (\ref{jump}) allows one to invert
formula (\ref{phi}) in terms of the Neumann jump
operator $\hat {\cal N}$,  the operator that takes a function
$f$ on $\Gamma$ to the jump of the normal derivative of its harmonic
continuations to the domains inside and outside $\Gamma$ (denoted by
$f_H$ and $f^H$ respectively): 
\beq\label{Neumann}
 {\cal N}f=\p_n^+f_H - \p_n^-f^H.
\eeq
See Appendix C for details. We will also use the Dirichlet-to-Neumann operators
$\hat {\cal N}^{\pm}$ which send a function $f$ on $\Gamma$ to normal derivatives of its
harmonic continuations to the interior (exterior) domains:
\beq\label{Neumann1}
\hat {\cal N}^+f=\p_n^{+}f_H, \quad \hat {\cal N}^- f=\p_n^{-}f^H,
\eeq
so that $\hat {\cal N}=\hat {\cal N}^{+}-\hat {\cal N}^{-}$.
In what follows, we denote the interior domain by ${\sf D}$, then the exterior one is
$\CC \setminus {\sf D}$, $\Gamma =\p {\sf D}$. 
Using
the Neumann jump operator, we write equation (\ref{jump})
as 
\beq\label{jump1}
\rho (z)=\frac{1}{2\pi}\, |w'_{\rm ext}(z)| + \frac{1}{4\pi \beta}\,
\hat {\cal N}\varphi (z),
\eeq
where $w_{{\rm ext}}$ is the conformal map from $\CC \setminus {\sf D}$
onto the exterior of the unit circle such that
$w_{{\rm ext}}(z ) = z/r + O(1)$ as $z\to \infty$ with a real positive $r$ which is called
the conformal radius of 
the exterior domain. Note that 
$|w'_{\rm ext}(z)|=\p_n ^{-}\log |w_{\rm ext}(z)|$, $z\in \Gamma$.

Correlation functions of densities can be obtained as
variational derivatives of $\log Z_N$ w.r.t. the potential.
For example,
\beq\label{var}
\lbracket \rho (z) \rbracket =\hbar \,
\frac{\delta \log Z_N}{\delta W(z)}\,,
\;\;\;\;
\lbracket \rho (z_1 )\rho (z_2)\rbracket _{{\rm c}}=\hbar \,
\frac{\delta \lbracket \rho (z_1)\rbracket }{\delta W(z_2)}=\hbar^2
\frac{\delta^2 \log Z_N}{\delta W(z_1) \delta W(z_2)}.
\eeq
Here $\lbracket \rho (z_1 )\rho (z_2)\rbracket _{{\rm c}}=
\lbracket \rho (z_1 )\rho (z_2)\rbracket -
\lbracket \rho (z_1) \rbracket \lbracket \rho (z_2) \rbracket$
is the connected part of the correlation function.
These formulas follow from the fact that
variation of the partition function w.r.t.
$W$ inserts $\sum_i \delta _{\Gamma}(z,z_i)$ into
the integral.
Clearly, 
\beq\label{meanrho}
\lbracket \rho (a) \rbracket =R(a), \quad
\lbracket \rho (a) \rho (b)\rbracket = R(a,b)+
\hbar \lbracket \rho (a) \rbracket \delta _{\Gamma}(a,b).
\eeq

Correlation functions of $\varphi$'s can be obtained from those
of $\rho$'s using equation (\ref{phi}).
An equivalent method
is to consider
linear response of the system to a small change of
the potential
$\delta_a W(z)= \varepsilon \log |z-a|^2$
which means inserting a point-like charge $\varepsilon \to 0$
at the point $a$. We have the exact relations
\beq\label{cf2}
\delta_a \log Z_N = -\frac{\varepsilon}{\beta \hbar}
\lbracket \varphi (a) \rbracket
\eeq
and
\beq\label{cf3}
\delta_a  \! \lbracket X \rbracket= -\frac{\varepsilon}{\beta \hbar}
\lbracket X \varphi (a) \rbracket_{{\rm c}}
\eeq
for any $X$.
In particular,
\beq\label{cf4}
\delta_a  \! \lbracket \varphi (z) \rbracket=
-\frac{\varepsilon}{\beta \hbar}
\lbracket \varphi (z) \varphi (a) \rbracket_{{\rm c}}.
\eeq

\subsection{Stress tensor}

A holomorphic traceless tensor 
\beq\label{T1}
T(z)=\beta ^2 \hbar^2 \sum_{j\neq k}
\frac{1}{(z-z_j)(z -z_k)} + \beta \hbar^2 \sum_j \frac{1}{(z-z_j)^2}
+2\beta \hbar ^2 \sum_j \frac{\p W(z_j)}{z-z_j}
\eeq
plays an important role. We refer to it as stress tensor and utilize it
in the next section. Here we express it in terms of 
the fields $\varphi$ and $\rho$:
\beq\label{T2}
T(z)=(\p \varphi (z))^2 +(1\! -\! \beta )\hbar \, \p^2 \varphi (z)
+2\beta \hbar \oint_{\Gamma} \frac{\p W(\xi )\rho (\xi )}{z-\xi }\, |d\xi |.
\eeq
The mean value $\lbracket T(z)\rbracket$ is a 
holomorphic function for all $z\notin \Gamma$ given by
\beq\label{T3}
\lbracket T(z)\rbracket =2\beta ^2 \oint_{\Gamma}
\frac{|d\xi |}{z-\xi}\oint_{\Gamma}
\frac{R(\xi ,\zeta )|d\zeta |}{\xi -\zeta} +\beta \hbar
\oint_{\Gamma}
\frac{R(\xi )|d\xi |}{(z-\xi )^2} +2\beta \hbar \oint_{\Gamma}
\frac{\p W(\xi )R(\xi )}{z-\xi}\, |d\xi|,
\eeq
where the identities (\ref{R3b}) are used.
On the contour $\Gamma$ it has a jump which follows from 
the Sokhotsky-Plemelj formula
$$
\oint_{\Gamma}\frac{g (\xi )|d\xi |}{\xi -z_{\pm}}=
\int_{\Gamma}\hspace{-13pt} {\bf -}\,
\frac{g (\xi )|d\xi |}{\xi -z}\, \pm \,
\pi \, \overline{\nu (z)}g (z),
$$
where $\nu (z)=-i dz /|dz|$ is the outward 
unit normal vector
to $\Gamma$ at the point $z$ and
$z_{\pm}$ means that $z$ approaches the contour from  the interior/exterior.
The second integral in (\ref{T3}) can be transformed as follows:
{\small
$$
\oint_{\Gamma}
\frac{R(\xi )|d\xi |}{(z-\xi )^2}=i\oint_{\Gamma}R(\xi )\overline{\nu (\xi )}\, 
d\frac{1}{\xi -z}=-i\oint_{\Gamma}\frac{d(R(\xi )\overline{\nu (\xi )})}{\xi -z}=
-\oint_{\Gamma}\frac{\bigl [i\p_s R(\xi ) +\kappa (\xi )R(\xi )\bigr ] \overline{\nu (\xi )}
\, |d\xi |}{\xi -z}.
$$
}
Here $\p_s$ is the tangential derivative along $\Gamma$ and
$\kappa (z)=i\nu (z)\p_s \overline{\nu (z)}$ is the curvature of the
contour at the point $z$. Using the notation
\beq\label{jump2}
[h (z)]_{\Gamma} = h(z_+) -h(z_-)\,, \quad \quad z\in \Gamma 
\eeq
introduced in Sec. 1
for the jump across $\Gamma$, we have:
\beq\label{T4}
\left [ \lbracket T(z)\rbracket \right ] _{\Gamma} =
2\pi \beta \, \overline{\nu (z)} \left (2\beta \oint_{\Gamma}
\frac{R(z, \zeta )|d\zeta |}{\zeta -z}- 2\hbar \p W(z)R(z)
-\hbar \, \overline{\nu (z)}\Bigl (\kappa (z) \! +\! i \p_s
\Bigr )R(z)\right ).
\eeq
The first integral is
convergent at $\zeta =z$ because $R(z,\zeta )\sim |z-\zeta |^{2\beta}$ as 
$\zeta \to z$.

In what follows it will be instructive to divide the mean value of $T$ in two parts,
``classical'' and ``quantum'': $\lbracket T\rbracket = 
\lbracket T\rbracket ^{(cl)}+\lbracket T\rbracket^{(q)}$.
The ``classical'' part is obtained from (\ref{T2}) by disregarding 
the connected part of the pair correlation function:
\beq\label{Tclas}
\lbracket T (z)\rbracket ^{(cl)}=
(\lbracket \p \varphi (z) \rbracket )^2 +(1\! -\! \beta )\hbar \, \p^2  \! \! 
\lbracket \varphi (z)\rbracket
+2\beta \hbar 
\oint_{\Gamma} \frac{\p W(\xi )\lbracket \rho (\xi )\rbracket }{z-\xi }\, |d\xi |.
\eeq
The ``quantum'' contribution 
\beq\label{Tquant}
\lbracket T(z)\rbracket^{(q)}=\lbracket (\p \varphi (z))^2 \rbracket _{\rm c}
\eeq
is the connected part.

\section{Exact relations at finite $N$}

There are two basic relations: one is an identity
for correlation functions which follow from the invariance
of the partition function under re-parametrizations of the
contour (the ``loop equation''), another describes the variation
of the partition function under small deformations of the contour.
As we shall see, both can be expressed in terms of the jump of $\lbracket T(z)\rbracket$
across the contour.

\subsection{Loop equation}

Let us start from
the obvious identity
\beq\label{rep1}
\sum_j \oint_{\Gamma}\! \ldots \!
\oint_{\Gamma}\p_{s_j}\left ( \epsilon (z_j)
\prod_{i<k}|z_i -z_k |^{2\beta}\prod_m e^{W(z_m)}
\right )\prod_{l=1}^{N}|dz_l| =0,
\eeq
which expresses invariance
of the partition function under re-parametrizations of the
contour (changes of the integration variables).
Here $\epsilon (z)$ is an arbitrary differentiable function on $\Gamma$.
In terms of the mean value (\ref{mean}) the identity takes the form 
$$
\lbracket
\sum_j \left (\p_{s_j}\epsilon (z_j) +
\epsilon (z_j)\p_{s_j}W(z_j)\right )+\beta \! \sum_{j\neq k}
\epsilon (z_j)\p_{s_j}\log |z_j \! -\! z_k |^2
\rbracket 
=0,
$$
which, according to (\ref{R3b}), can be rewritten through the
distribution functions:
\beq\label{rep2}
\oint_{\Gamma}|dz|\epsilon (z)\left (
\hbar \p_s W(z)R(z)-\hbar \p_s R(z)+\beta \oint_{\Gamma}
R(z, \xi )\p_s \log |z-\xi |^2 |d\xi | \right ) =0.
\eeq
Since this identity holds for any
$\epsilon (z)$, it implies that
\beq\label{chain1}
\beta \! \oint _{\Gamma} R(z,\xi )
\p_s \log |z-\xi |^2 \,
|d \xi | +\hbar \p _s W(z) R(z)-\hbar \p _s R(z)=0.
\eeq
In fact this exact relation is 
the first equation of an infinite system connecting
multi-point distribution functions,
 called
the BBGKY hierarchy (after Bogoliubov, Born, Green, Kirkwood and Yvone).
Note also that equation (\ref{chain1})
can be obtained in a simpler way by differentiating the definition of
$R(z)$ (\ref{R1}) with respect to $z$ along the contour $\Gamma$
(cf. \cite{Landafshitz}).

It remains to notice that equation (\ref{chain1}) is equivalent to
the following condition for the jump of $\lbracket T(z)\rbracket$
given by (\ref{T4}):
\beq\label{loop2}
{\cal I}m \left (\nu^2 (z)[ \lbracket T(z)\rbracket ]_{\Gamma}\right )=0\,,
\quad z\in \Gamma .
\eeq
In the frame with axes 
normal and tangential to the contour   
the components of the ``stress  tensor'' $T$ read
\beq\label{var8}
\begin{array}{l}
4T_{nn}=-4T_{ss}=\, \nu^2 T +\bar \nu^2 \bar T,
\\ \\
4T_{sn}=i\nu^2 T -i\bar \nu^2 \bar T.
\end{array}
\eeq
Then
the loop equation states that the tangential-normal component
is continuous across the boundary:
\beq\label{loop2a}
[ \lbracket T_{sn}(z) \rbracket ]_{\Gamma}  =0,
\quad z\in \Gamma .
\eeq
We comment that the boundary condition is different from that used in the 
boundary conformal field theory. There $\lbracket T_{sn}(z)\rbracket =0$ 
when the argument approaches
the boundary from one side.

If $\Gamma$ is the unit circle, then $\nu (z)=z$. Equation (\ref{chain1}) acquires the form
$$
\beta \oint_{|\xi |=1}R(z, \xi )\frac{z+\xi}{z-\xi}\, \frac{d\xi}{\xi}+
\hbar \p_s W(z)R(z)-\hbar \p_s R(z)=0.
$$
It is equivalent to the 
more customary form of the loop equation
{\small
$$
\sum_j \left < \beta \sum_{k\neq j}\frac{z}{(z-z_j)(z-z_k)}+
\frac{z}{(z-z_j)^2}+\frac{z_j\p W(z_j)\! -\! \bar z_j \bar \p W(z_j)}{z-z_j}-
\frac{\beta N \! -\! \beta \! +\! 1}{z-z_j}
\right >=0
$$
} 
which follows from the obvious identity
$$
\sum_j \oint_{\Gamma}\! \ldots \!
\oint_{\Gamma}\p_{s_j}\left ( \frac{1}{z-z_j}
\prod_{i<k}|z_i -z_k |^{2\beta}\prod_m e^{W(z_m)}
\right )\prod_{l=1}^{N}|dz_l| =0
$$
valid for any $z\in \CC$ (outside of  the unit circle).

Using the expression (\ref{T2}) for the stress tensor, we rewrite 
(\ref{loop2a}) in terms of correlation
functions of the fields $\varphi$ and $\rho$. We have:
$$
{\cal I}m \left \{\nu^2 (z)\left (
\Bigl [\left < (\p \varphi (z))^2 \right >\Bigr ]_{\Gamma} +
\hbar (1\! -\! \beta )\Bigl [\left <\p^2 \varphi (z)\right >\Bigr ]_{\Gamma}
+2\beta 
\hbar \left [\oint_{\Gamma}\frac{\p W(\xi )\left < \rho (\xi )
\right >}{z-\xi }\, |d\xi |\right ]_{\Gamma}
\right ) \right \}=0.
$$
Let us consider the three terms in the l.h.s. separately. 
In the first term, we separate the 
connected part and use equation (\ref{ap2}) from Appendix A to write the other two:
$$
2{\cal I}m \left \{\nu^2 \Bigl [\left < (\p \varphi )^2 \right >\Bigr ]_{\Gamma}\right \}=
-\left [ \p_n \! \left < \varphi \right >\right ]_{\Gamma}\p_s \!
\left < \varphi \right >+
2{\cal I}m \left \{\nu^2 \Bigl [\left < (\p \varphi )^2 \right >_{\rm c}\Bigr ]_{\Gamma}\right \}.
$$
Here and below in similar calculations we take into account that 
the function $\left <\varphi \right >$ is continuous 
across the contour, so the jumps like $[\p_s \! \left <\varphi \right >]_{\Gamma}$ vanish.
In the second term, we use equation (\ref{ap4}) from Appendix A to get
$$
2{\cal I}m \left \{\nu^2 \Bigl [\left <\p^2 \varphi \right >\Bigr ]_{\Gamma}\right \}=
-\p_s \left [\p_n \! \left < \varphi \right >\right ]_{\Gamma}
=-4\pi \beta \p_s \left < \rho \right >.
$$
In the third term we apply the Sokhotsky-Plemelj formula:
$$
2{\cal I}m \left \{\nu^2 (z)\left [
\oint_{\Gamma}\frac{\p W(\xi )\left < \rho (\xi )\right >}{z-\xi }\, |d\xi |\right ]_{\Gamma}
\right \}=2\pi \p_s W(z) \! \left < \rho (z) \right >.
$$
Combining the three terms, we bring the loop equation to the  form
\beq\label{it1}
\begin{array}{l}
\lbracket \rho (z) \rbracket  \Bigl (\p_s \! \lbracket 
\varphi (z)\rbracket - \hbar \p_s W(z) \Bigr )
-(\beta \! -\! 1) \hbar \, \p_s \! \lbracket \rho (z) \rbracket
\\ \\
\hspace{1cm}=\,\, \displaystyle{\frac{1}{2\pi \beta}\,
{\cal I}m \! \left ( \nu^2 (z) \left [ \,
\lbracket (\p \varphi (z))^2 \rbracket _{{\rm c}}
\, \right ]_{\Gamma}\right )}
\end{array}
\eeq
which will be used for the large $N$ expansion.
The l.h.s. comes from the  classical part of the 
stress-energy tensor while the r.h.s. is due to the ``quantum''
fluctuations. Note that the three terms of the loop equation (\ref{it1}) have orders
$O(\hbar^0)$, $O(\hbar^1)$ and $O(\hbar^2)$ respectively. 

A comment on the relation to conformal field theory (CFT) is in order. 
The boundary condition there is $\lbracket T_{sn}\rbracket =0$, not 
$[\lbracket T_{sn}\rbracket ]_{\Gamma}=0$.  
With this boundary condition, under a change of the geometry 
(a deformation of the contour) correlation 
functions transform according to the CFT rules. This is not the case in our model,
and so the similarity with CFT is not complete.

\subsection{Deformations of  contour}

Let us describe small deformations of the contour $\Gamma$
by a normal displacement $\delta n(z)$ of the point $z \in \Gamma$
(positive for displacements in the outward direction).
Below we use the rule for variation of the contour
integral $I=\oint_{\Gamma}f (z)|dz|$ given in \cite{WZ03} (see also Appendix E). 
Let $f$ be a function
defined in a strip-like neighborhood of the contour, 
then
\beq\label{varI}
\delta I = 
\oint_{\Gamma}\delta n (z)(\p_n \! +\! \kappa (z))f(z) |dz|
+\oint_{\Gamma}\delta f (z)|dz|.
\eeq
Here $\kappa (z)$ is the curvature of $\Gamma$ at the point $z$.
The first term is due to the deformation of the contour.
The second contribution is due to the variation of the function $f$ itself (in the case
if it depends on the shape of the contour).

According to (\ref{varI}),
we have:
$$
\delta Z_N = \sum_j \oint_{\Gamma} \! \ldots \!
\oint_{\Gamma}\delta n(z_j)
\Bigl (\p_{n_j} \! +\! \kappa (z_j)\Bigr )
\left (\prod_{i<k}|z_i -z_k |^{2\beta}\prod_m e^{W(z_m)}
\right )
\prod_{l=1}^{N}|dz_l|.
$$
Similarly to the derivation of equation (\ref{rep2}) from identity
(\ref{rep1}), we write:
$$
\delta  \log Z_N =\lbracket \sum_j
\Bigl ( \kappa (z_j) +\p_{n_j}W(z_j) +
\beta \sum_{k\neq j} \p_{n_j}\log |z_j - z_k|^2 \Bigr )
\delta n(z_j)\rbracket
$$
or, in terms of the mean density and the pair distribution function,
$$
\hbar^2 \delta \log Z_N =
\oint_{\Gamma}
|dz|\delta n(z) \! \left (\hbar \p_n W(z)R(z) + \hbar \kappa (z)R(z)+
\beta \! \oint_{\Gamma} \! R(z, \xi )\p_n \log |z-\xi |^2 |d\xi |\right ).
$$
Comparing with (\ref{T4}), one can rewrite this relation
in terms of the jump of $\lbracket T(z)\rbracket$:
\beq\label{jump102}
\hbar^2 \delta \log Z_N =
-\, \frac{1}{2\pi \beta}\oint_{\Gamma}
{\cal R}e \! \left (\nu^2 (z) \Bigl [\lbracket T(z)
\rbracket \Bigr ]_{\Gamma} \right )
\delta n (z) \, |dz|.
\eeq
It can be written in some other forms. In terms of the normal-normal component 
of the stress-energy tensor we have
\beq\label{nn}
\hbar^2 \delta \log Z_N =-\frac{1}{\pi \beta}
\oint_{\Gamma} \Bigl [ \left < T_{nn}(z)\right >\Bigr ]_{\Gamma} \delta n (z) \, |dz|.
\eeq
Combining (\ref{jump102}) with the loop equation which says that the
quantity $\Bigl < \nu^2 (z) [T(z)]_{\Gamma} \Bigr >$
is real, we can also write
\beq\label{jump102a}
\hbar^2 \delta \log Z_N =
-\, \frac{1}{2\pi \beta}\oint_{\Gamma}
\nu^2 (z) \, \Bigl <[T(z)]_{\Gamma} \Bigr >\,
\delta n (z)  |dz|.
\eeq

For some particular deformations equation (\ref{jump102})
gets simplified. For example, set $\delta n(z)=\delta _a n (z )$, where
$$
\delta _a n (z )=2\varepsilon \, {\cal R}e  \left (
\frac{\overline{\nu (z)}}{z -a}\right )
$$
then a simple calculation of residues yields
\beq\label{var6}
\beta \hbar^2 \delta \log Z_N =-2\varepsilon \,
{\cal R}e \lbracket T(a)\rbracket .
\eeq

For the variation $\delta_{W} \log Z_N$ caused by a small change of the potential, we obtain:
\beq\label{var2}
\hbar \delta_{W} \log Z_N =\left < \hbar \sum_j \delta W(z_j)\right >=
\oint_{\Gamma}\left < \rho (z)\right > \delta W(z)|dz|.
\eeq
The total variation is $\delta \log Z_N + \delta_{W} \log Z_N$.

Let us represent the jump of $\left < T_{nn}\right >$ in the form similar to (\ref{it1}).
We have:
$$ 
4\Bigl [ \left < T_{nn}(z)\right > \Bigr ]_{\Gamma}
=\, 2{\cal R}e \left \{\nu^2 (z)\left (
\vphantom{\oint_{\Gamma}\frac{\p W(\xi )\left < \rho (\xi )\right >}{z-\xi }}
\Bigl [\left < (\p \varphi (z))^2 \right >\Bigr ]_{\Gamma} +
\hbar (1\! -\! \beta )\Bigl [\left <\p^2 \varphi (z)\right >\Bigr ]_{\Gamma}\right. \right.
$$
$$
\phantom{aaaaaaaaaaaaa}\left. \left.
+2\beta 
\hbar \left [\oint_{\Gamma}\frac{\p W(\xi )\left < \rho (\xi )\right >}{z-\xi }\, 
|d\xi |\right ]_{\Gamma}
\right ) \right \}.
$$
In the first term, we separate the 
connected part and use equation (\ref{ap1}) from Appendix A in the rest:
$$
2{\cal R}e \left \{\nu^2 \Bigl [\left < (\p \varphi )^2 \right >\Bigr ]_{\Gamma}\right \}=
\frac{1}{2} \left [ \p_n \! \left < \varphi \right >\right ]_{\Gamma}\!
\Bigl ( \p_n^+ \! \left < \varphi \right >+\p_n^- \! \left < \varphi \right >\Bigr)+
2{\cal R}e \left \{\nu^2 \Bigl [\left < (\p \varphi )^2 
\right >_{\rm c}\Bigr ]_{\Gamma}\right \}.
$$
In the second term, we use equation (\ref{ap3}) from Appendix A to get
$$
2{\cal R}e \left \{\nu^2 \Bigl [\left <\p^2 \varphi \right >\Bigr ]_{\Gamma}\right \}=
-\kappa \left [\p_n \! \left < \varphi \right >\right ]_{\Gamma}
=-4\pi \beta \kappa \lbracket \rho \rbracket .
$$
In the third term we apply the Sokhotsky-Plemelj formula:
$$
2{\cal R}e \left \{\nu^2 (z)\left [
\oint_{\Gamma}\frac{\p W(\xi )\left < \rho (\xi )\right >}{z-\xi }\, |d\xi |\right ]_{\Gamma}
\right \}=-2\pi \p_n W(z) \! \left < \rho (z) \right >.
$$
Combining the three terms, we obtain:
\beq\label{tnn}
\begin{array}{lll}
4\Bigl [ \left < T_{nn}(z)\right > \Bigr ]_{\Gamma}&=&
2\pi \beta \left < \rho (z)\right >\Bigl \{
\p_n^+ \! \left <\varphi (z)\right > +\p_n^- \! \left <\varphi (z)\right >-2\hbar \p_n W(z)+
2(\beta -1)\hbar \kappa (z)\Bigr \}
\\ && \\
&&+\,\,\, 2{\cal R}e \left \{\nu^2 (z)\Bigl [\left < (\p \varphi (z))^2 
\right >_{\rm c}\Bigr ]_{\Gamma}\right \}.
\end{array}
\eeq
Similarly to (\ref{it1}), the first line comes from the classical part of the 
stress tensor while the second one is due to the quantum 
contribution.

\section{Large $N$ expansion}

The large $N$
(small $\hbar$) behavior
of $\log Z_N$ is of the form (\ref{large1}). 
Our task is to obtain explicit expressions for the contributions to the free energy
$F_0$, $F_1$ and $F_2$. 

We assume that $\lbracket \rho (z) \rbracket$ can be expanded
in integer powers of $\hbar$:
\beq\label{expan1}
\lbracket \rho (z) \rbracket = \rho_0 (z)+\hbar \rho_{1}(z)+
\hbar^2 \rho_2(z) +\ldots
\eeq
Then a similar expansion holds for $\lbracket \varphi (z) 
\rbracket$:  $\lbracket \varphi  \rbracket =
\varphi_0 + \hbar \varphi_{1}+\hbar^2 \varphi_{2} + \ldots \, $. 
Their coefficients are related as
$4\pi \beta \rho _j =[\p_n \varphi _j]_{\Gamma}$.

\subsection{The leading order}

In the leading order, the problem is equivalent to a 2D electrostatic
problem of finding the equilibrium distribution of a charged conducting fluid.
It can be easily solved by requiring the total
electrostatic potential created by the charges and by the background to be
constant along the contour. This will be the leading order of the regular iterative 
procedure based on the loop equation we develop below.

The connected correlation function in the right hand side of the loop equation (\ref{it1}) 
is of order $O(\hbar^2)$. Therefore, 
at the 0th step, the loop equation
states
that
$$
\rho_0 \, \p_s \varphi _0 =0
$$
which means that $\varphi _0$ is constant along
$\Gamma$.
Then 
\beq\label{lead1}
\rho_0 (z)= \frac{1}{2\pi}\,|w'_{{\rm ext}}(z)| , \quad z\in \Gamma .
\eeq
This is the solution to the electrostatic problem of continuous charge distribution
over the contour. In this approximation discreetness of the particles is neglected.
The charges form a simple layer whose electric potential
$\varphi _0 (z)=-2\beta \oint_{\Gamma}
\log |z-\xi |\rho_0 (\xi )|d\xi |$ reads
\beq\label{lead100}
\varphi _0 (z)=\left \{
\begin{array}{ll}
-2\beta  \log r 
&\quad \mbox{for $z\in {\sf D}$}
\\ &\\
-2\beta \log r -2\beta  \log |w_{{\rm ext}}(z)|
&\quad \mbox{for $z\in \CC \setminus {\sf D}$.}
\end{array} \right.
\eeq
Next, we use (\ref{tnn}) to find that
\beq\label{lead101}
4\Bigl [ \left < T_{nn}\right > \Bigr ]_{\Gamma}\cong 
2\pi \beta \rho_0 \left (\p_{n}^{+}\varphi_0 +\p_{n}^{-}\varphi_0 \right ).
\eeq
Here and below $\cong$ means equality in the leading order (up to higher powers of $\hbar$). 
This should be substituted into 
(\ref{nn}). After some simple transformations, one can find $\delta F_0$ with $F_0$ given by 
\beq\label{lead5}
F_0 = \beta  \log r .
\eeq 
This is the electrostatic energy of the charge distribution with density (\ref{lead1}).

Two-point correlation functions in the leading order can be found using the
variational formulas (\ref{var}) or simple electrostatic arguments.
Explicitly, the result is:
\beq\label{lead6}
\frac{\lbracket \varphi (z)\varphi (\zeta )
\rbracket _{{\rm c}}}{2\beta \hbar^2}\cong
\left \{
\begin{array}{l}
\displaystyle{
G_{{\rm ext}}(z, \zeta )\! -\! G_{{\rm ext}}(z, \infty )\! -\!
G_{{\rm ext}}(\infty , \zeta )\! -\! \log \frac{|z-\zeta |}{r}\,,
\quad z,\zeta \in \Dc}
\\ \\
\displaystyle{
G_{{\rm int}}(z, \zeta ) - \log \frac{|z-\zeta |}{r}\,,
\quad \quad z,\zeta \in \DD}
\\ \\
\displaystyle{
-G_{{\rm ext}}(\infty , \zeta )- \log \frac{|z-\zeta |}{r}\,,
\quad \quad z\in \DD , \zeta \in \Dc}.
\end{array}
\right.
\eeq
Here $G_{\rm int}$, $G_{\rm ext}$ are Green's functions of the Dirichlet boundary problems
for the Laplace operator in the interior and exterior domains respectively (see Appendix B).
For derivatives of the fields this gives:
\beq\label{lead6a}
\frac{\lbracket \p \varphi (z)\p \varphi (\zeta )
\rbracket _{{\rm c}}}{\beta \hbar^2}\cong
\left \{
\begin{array}{l}
\displaystyle{
\frac{w'_{\rm ext}(z)w'_{\rm ext}(\zeta )}{(w_{\rm ext}(z)-w_{\rm ext}(\zeta ))^2}-
\frac{1}{(z-\zeta )^2}\,,
\quad z,\zeta \in \Dc}
\\ \\
\displaystyle{
\frac{w'_{\rm int}(z)w'_{\rm int}(\zeta )}{(w_{\rm int}(z)-w_{\rm int}(\zeta ))^2}-
\frac{1}{(z-\zeta )^2}\,,
\quad \quad z,\zeta \in \DD}
\\ \\
\displaystyle{
-\, \frac{1}{(z-\zeta )^2} \,,
\quad \quad z\in \DD , \zeta \in \Dc}.
\end{array}
\right.
\eeq
At merging points
$$
\lim_{\zeta \to z}\left (
\frac{w'(z)w'(\zeta )}{(w(z)-w(\zeta ))^2} -\frac{1}{(z-\zeta )^2}\right )=
\frac{1}{6}\{ w; z\},
$$
where
$$
\{ w; z\}=\frac{w'''(z)}{w'(z)}-\frac{3}{2}\left (
\frac{w''(z)}{w'(z)}\right )^2 =\p^2 \log w' - \frac{1}{2}
(\p \log w' )^2
$$
is the Schwarzian derivative.
Then we obtain from (\ref{lead6a}):
\beq\label{lead7}
\lbracket (\p \varphi (z))^2\rbracket _{{\rm c}}\cong
\left \{
\begin{array}{l}
\displaystyle{
\frac{\beta \hbar^2}{6}\, \{w_{{\rm ext}}; z\}\,,
\quad \quad z\in \Dc}
\\ \\
\displaystyle{
\frac{\beta \hbar^2}{6}\, \{w_{{\rm int}}; z\}\,, \,\,\,
\quad \quad z\in \DD}.
\end{array}\right.
\eeq

\subsection{General scheme of iterative solution}

We start from the loop equation in the form (\ref{it1}). 
According to (\ref{lead7}), the r.h.s.
is a priori expected to be of the order $O(\hbar^2)$.
However, 
the quantity ${\cal I}m \left ( \nu^2 (z)\{w;z\} \right )$ is the 
tangential derivative of the curvature $\kappa(s)$ which is continuous across the boundary:
\beq\label{it2}
\p_s \kappa (s)={\cal I}m \left ( \nu^2 (z)\{
w_{{\rm int}};z\} \right )=
{\cal I}m \left ( \nu^2 (z)\{
w_{{\rm ext}};z\} \right ).
\eeq
Therefore, the r.h.s. 
is actually at least $O(\hbar^3)$ and the loop equation
simplifies to
\beq\label{it3}
\hbar \p_s W(z)-
\p_s \! \lbracket \varphi (z)\rbracket
+(\beta \! -\! 1) \hbar \, \p_s  \log \lbracket \rho (z) \rbracket=O(\hbar^3).
\eeq
Integrating, we write it as
\beq\label{it3a}
\lbracket \varphi (z)\rbracket =
\hbar W (z)+ \hbar (\beta \! -\! 1)  \log \lbracket \rho (z) \rbracket
+\, \lambda \, + \, O(\hbar^3),
\eeq
where the constant $\lambda$ reads
\beq\label{harm2}
\lambda =-2\beta  \log r-\hbar W^H(\infty )-
\hbar (\beta \! -\! 1)(\log \lbracket \rho \rbracket )^H (\infty ).
\eeq
It can be found by the harmonic extension of (\ref{it3a}) to the exterior of the contour.
The limit $z\to \infty$ determines the constant.

Finally, applying $\hat {\cal N}$ to equation (\ref{it3a}), we arrive at the equation
\beq\label{it4}
\lbracket \rho (z) \rbracket -\rho_0 (z) =\frac{\hbar}{4\pi \beta}\, \hat {\cal N}
W(z)+
\frac{(\beta \! -\! 1)\hbar}{4\pi \beta} \, \hat {\cal N}\log
\lbracket \rho (z) \rbracket +\, O(\hbar^3),
\eeq
where $\rho_0$ is given by (\ref{lead1}). 
The iterations give
\beq\label{it6}
\begin{array}{l}
\displaystyle{\rho_{1}=\frac{\beta \! -\! 1}{4\pi \beta}\,
\hat {\cal N}\log \rho_0}+ \frac{1}{4\pi \beta}\, \hat {\cal N}W
\\ \\
\displaystyle{\rho_{2}=
\left (\frac{\beta \! -\! 1}{4\pi \beta}\right )^2
\hat {\cal N}\rho_0^{-1}\hat {\cal N}\, \log \rho_0 +
\frac{\beta \! -\! 1}{(4\pi \beta )^2}\hat {\cal N}\rho_0^{-1}\hat {\cal N}\,W.}
\end{array}
\eeq

The $\hbar$-expansion of $\lbracket \rho \rbracket$ implies
an $\hbar$-expansion of the mean value of the stress tensor
$\lbracket T(z) \rbracket$:
$\lbracket T\rbracket =T_0 + \hbar T_{1}+ \hbar^2 T_{2}+\ldots \, $.
According to (\ref{T2}),
the first few terms of the latter are:
\beq\label{it9}
\begin{array}{l}
T_0  = (\p \varphi _0)^2 
\\ \\
\displaystyle{
T_{1} = 2 \p \varphi _0 \p
\varphi_{1}+ (1\! -\! \beta )\p^2 \varphi_0 +2\beta
\oint_{\gamma}\frac{\p W(\xi )\rho_0(\xi )}{z-\xi}\, |d\xi |}
\\ \\
\displaystyle{
T_2  =(\p \varphi _{1})^2+
(1\! -\! \beta )\p^2 \varphi_{1}
+2 \p \varphi _0 \p \varphi_{2} +
2\beta \oint_{\Gamma}\frac{\p W(\xi )\rho_1(\xi )}{z-\xi}\, |d\xi |
+\lim_{\hbar \to 0}
\hbar^{-2}\lbracket (\p \varphi )^2 \rbracket _{{\rm c}}.}
\end{array}
\eeq
A possible iterative procedure is as follows:
first use the ``loop equation''
$
\Bigl [( T_{sn})_j \Bigr ]_{\Gamma}=0
$
to find $\varphi _j$ on $\Gamma$,
then extend  $\varphi _j$ harmonically to the complex plane (which is necessary 
for finding normal derivatives) and
find $F_j$ from the variation 
\beq\label{it8}
\delta F_j =-\, \frac{1}{\pi \beta}
\oint_{\Gamma} \Bigl [ ( T_{nn})_j \Bigr ]_{\Gamma}
\delta n ds \, + \, \oint_{\Gamma}\rho_{j-1} \delta W ds.
\eeq
Here $\delta W$ is a change of the potential $W$ induced by a small deformation 
of the contour in the case when $W$ explicitly depends
on the contour (below we discuss such example).

For the $\hbar$-expansion of 
$\Bigl [ \left < T_{nn}\right > \Bigr ]_{\Gamma} $ one can also use 
equation (\ref{tnn}).
It is convenient to express the r.h.s. in terms of $\lbracket \rho \rbracket$.
Regrouping the terms in (\ref{tnn}) and
substituting the normal derivatives $\p_n^{\pm}\lbracket \varphi \rbracket$, we obtain:
\beq\label{tnn1}
\begin{array}{lll}
4\Bigl [ \left < T_{nn}\right > \Bigr ]_{\Gamma}&=&
-\, 4\pi \beta  \lbracket \rho \rbracket \Bigl ( \beta |w'_{\rm ext}|
+\hbar \p_n (W-\frac{1}{2}(W_H +W^H)\Bigr )
\\ && \\
&&+\,\, 4\pi \beta (\beta \! -\! 1)\hbar \lbracket \rho \rbracket
\Bigl ( \frac{1}{2}\p_n ((\log \lbracket \rho \rbracket )_H +(\log \lbracket \rho \rbracket )^H)
+\kappa \Bigr )
\\ && \\
&&+\,\,\, 2{\cal R}e \left \{\nu^2 \Bigl [\left < (\p \varphi )^2 
\right >_{\rm c}\Bigr ]_{\Gamma}\right \}.
\end{array}
\eeq
The first line is $O(1)$, the second one is $O(\hbar )$ and the third one is $O(\hbar^2)$.

\subsection{The next-to-leading order}

The correction $\rho_1$ is given by (\ref{it6}). 
The corresponding correction to $\lbracket \varphi \rbracket$ is given by
\beq\label{next1}
\varphi _{1}(z)=(\beta -1) \left ( \log \rho_0(z)-
(\log \rho_0 )^H (\infty )\right ) +W (z) -W^H(\infty), 
\quad z\in \Gamma .
\eeq
The constant term is found from the requirement that
$\varphi _{1}(z)\to 0$ as $|z|\to \infty$.
The values of $\varphi_{1}$ outside the contour are obtained by the harmonic continuations.
Further, using (\ref{tnn1}), we find:
\beq\label{tnn2}
4\Bigl [ (T_{nn})_{{}_1} \Bigr ]_{\Gamma}=2\beta |w'_{\rm ext}|\Bigl ( (\beta -1)
\Bigl (\p_n^- (\log \rho_0)^H \! +\! \kappa \Bigr )
+\p_n^- (W^{H}\! -\! W)\Bigr ).
\eeq
This is
the variation of the following simple
functional:
\beq\label{next3}
F_{1} = (\beta \! -\! 1)
\oint_{\Gamma} \rho_0 \log \rho_0 \, ds +
\oint_{\Gamma} \rho_0 W \, ds +c_1.
\eeq

The first term is the ``entropy-like'' contribution. It is due to entropy and excluded volume.
The integral can be easily calculated and the result is expressed through 
the logarithm of the conformal
radius $r$ of the exterior domain:
\beq\label{next3a}
F_1=-(\beta \! -\! 1)\log (2\pi r)+\oint_{\Gamma} \rho_0 W \, ds +c_1.
\eeq
The constant $c_1$ does not depend on the shape of the contour and can not be determined
from the variation. It can be found from a comparison with the exact result for the circle
(see below).

\subsection{Calculation of $F_2$}

First of all, let us separate the classical and quantum parts in $F_2$
according to (\ref{Tclas}), (\ref{Tquant}):
$F_2 = F_{2}^{(cl)}+F_{2}^{(q)}$. 
Their variations are determined by the respected
parts of $T_{2}$: 
$$
\begin{array}{l}
T_{2}^{(cl)} =
(\p \varphi _{1})^2+
(1\! -\! \beta )\p^2 \varphi_{1}
+2 \p \varphi _0 \p \varphi_{2}+\mbox{the $W$-dependent part},
\\ \\
T_{2}^{(q)} =\hbar^{-2}\!
\lbracket (\p \varphi )^2 \rbracket _{{\rm c}}.
\end{array}
$$
Note that the previously found 
$F_0$ and $F_1$ are in this sense  classical quantities.
The order $\hbar^2$ is the first one when the 
quantum correction appears.

\paragraph{The classical part.}
The direct expansion of (\ref{tnn1}) to the order $\hbar^2$ gives
\beq\label{TTT}
\begin{array}{lll}
4\Bigl [ (T^{(cl)}_{nn})_{{}_2}\Bigr ]_{\Gamma}&=&
-4\pi \beta^2 |w'_{\rm ext}|\rho_2 +
2\pi \beta \rho_1
\Bigl (\p_n^+ (W_{H}-W)+\p_n^- (W^{H}-W)\Bigr )
\\ && \\
&& \,\, +\,\, 2\pi \beta  (\beta \! -\! 1) \rho_1 
\Bigl (\p_n^+ (\log \rho_0)_H +\p_n^- (\log \rho_0)^H +2\kappa \Bigr )
\\ && \\
&& \,\, +\,\, 2\pi \beta  (\beta \! -\! 1) \rho_0
\Bigl (\p_n^+ (\rho_1/\rho_0)_H +\p_n^- (\rho_1/\rho_0)^H \Bigr ).
\end{array}
\eeq
The final result for $F_{2}^{(cl)}$ is:
\beq\label{F2cl}
F_{2}^{(cl)}=\frac{1}{8\pi \beta}\oint_{\Gamma}
\Bigl ((\beta \! -\! 1)\log \rho_0 +W \Bigr )\hat {\cal N}
\Bigl ((\beta \! -\! 1)\log \rho_0 +W \Bigr )ds.
\eeq
One can verify that the variation of this functional under small deformations 
of the contour is given by (\ref{it8}) with $T_{nn}$ as in (\ref{TTT}).

\paragraph{The quantum part.}
Using (\ref{lead7}) it is straightforward to find
the variation of the quantum part:
\beq\label{it12}
\delta F_{2}^{(q)}= \frac{1}{12\pi}
\oint _{\Gamma}{\cal R}e \Bigl ( \nu ^2 (z)\Bigl (
\{w_{{\rm ext}};z\} \! -\! \{w_{{\rm int}};z\}\Bigr )\Bigr )
\delta n(z) |dz|.
\eeq
In order to understand this result, 
we should compare it with the variations of the
regularized determinants of the Laplace operators $\Delta =4\p \bar \p$ in the interior
and in the exterior domains ${\sf D}$, $\CC \setminus {\sf D}$ with Dirichlet
boundary conditions. These determinants are given by formulas of the Polyakov-Alvarez type
\cite{Polyakov,Alvarez}, see also \cite{VDOP82,ADFO86,OPS88,HZ99} and Appendix E. 
For planar domains
they are given by integrals over boundary:
\beq\label{pa1}
\begin{array}{lll}
\log \det \left ( -\Delta _{{\rm int}}\right )&=&-\, \displaystyle{
\frac{1}{12\pi} \oint_{\Gamma}\left ( \psi_{\rm int}\p_n \psi_{\rm int} -
2\psi_{\rm int}e^{\psi_{\rm int}}\right )ds}
\\ && \\
&=& \displaystyle{
\frac{1}{12\pi} \oint_{\Gamma}\Bigl ( \psi_{\rm int}\p_n \psi_{\rm int} +
2\kappa \psi_{\rm int}\Bigr )ds},
\end{array}
\eeq
\beq\label{pa1a}
\begin{array}{lll}
\log \det \left ( -\Delta _{{\rm ext}}\right )&=& \displaystyle{
\frac{1}{12\pi} \oint_{\Gamma}\left ( \psi_{\rm ext}\p_n \psi_{\rm ext} -
2\psi_{\rm ext}e^{\psi_{\rm ext}}\right )ds}
\\ && \\
&=& \displaystyle{
-\, \frac{1}{12\pi} \oint_{\Gamma}\Bigl ( \psi_{\rm ext}\p_n \psi_{\rm ext} +
2\kappa \psi_{\rm ext}\Bigr )ds},
\end{array}
\eeq
where $\psi_{\rm ext}(z)=\log |w'_{\rm ext}(z)|$, 
$\psi_{\rm int}(z)=\log |w'_{\rm int}(z)|$. Here $w_{\rm int}(z)$ is the conformal map
from ${\sf D}$ to the unit disk. We note that formulas 
for logarithms of determinants of this type should be understood 
modulo some additive normalization constants in the r.h.s. which do not depend on the shape of 
$\Gamma$. Hereafter, we omit these constants for simplicity. 
Equivalent expressions given by integrals over the unit circle in the $w$-plane 
in terms of the inverse conformal maps
$z_{\rm int}(w)$, $z_{\rm ext}(w)$ are more customary:
\beq\label{q1}
\log \det \left ( -\Delta _{{\rm int}}\right )=\, - \,
\frac{1}{12\pi} \oint_{|w|=1} \left (\phi _{{\rm int}}
\p_n \phi _{{\rm int}} +2 \phi _{{\rm int}}
\right )|dw|,
\eeq
\beq\label{q1a}
\log \det \left ( -\Delta _{{\rm ext}} \right )\,\, =\,\,\,
\frac{1}{12\pi} \oint_{|w|=1} \left (\phi _{{\rm ext}}
\p_n \phi _{{\rm ext}} +2 \phi _{{\rm ext}}
\right )|dw|,
\eeq
where $\phi_{{\rm int}} (w)=\log |z'_{{\rm int}}(w)|$,
$\phi_{{\rm ext}} (w)=\log |z'_{{\rm ext}}(w)|$.
A direct calculation using the rules explained in Appendix D shows that
\beq\label{q2}
\delta \log \det \left ( -\Delta _{{\rm int}}\right )\, =\,\,
\frac{1}{6\pi}\oint _{\Gamma}
\Bigl ( {\cal R}e \Bigl ( \nu ^2 \{w_{{\rm int}};z\}\Bigr )
-\kappa ^2 \Bigr ) \delta n  |dz|,
\eeq
\beq\label{q2a}
\delta \log \det
\left ( -\Delta _{{\rm ext}} \right )
=-\,  \frac{1}{6\pi}\oint _{\Gamma}
\Bigl ( {\cal R}e \Bigl ( \nu ^2 \{w_{{\rm ext}};z\}\Bigr )
-\kappa ^2 \Bigr ) \delta n  |dz|.
\eeq
Hence we can write, taking into account (\ref{it12}):
\beq\label{q3}
\delta F_{2}^{(q)}= -\frac{1}{2}\, \delta
\log \det
\left ( \Delta _{{\rm int}} \Delta _{{\rm ext}} \right )
\eeq
or
\beq\label{q3a}
F_{2}^{(q)}= -\frac{1}{2}
\log \det
\left ( -\Delta _{{\rm int}} \right )
-\frac{1}{2}
\log \det
\left (-\Delta _{{\rm ext}} \right )+\mbox{const},
\eeq
where the additive constant does not depend on the shape of the contour.

This result can be represented in yet another instructive form if one recalls
the surgery (Mayer-Vietoris type) formula for the functional
determinants of Laplace operators in 
complimentary domains \cite{HZ99,BFK92}:
\beq\label{surg}
\log \det (-\Delta_{\rm int})+\log \det (-\Delta_{\rm ext})+\log {\det} ' \hat {\cal N}
=\log P +\mbox{const},
\eeq
where $P=\oint_{\Gamma}ds$ is the perimeter of the boundary of the 
domain ${\sf D}$ (the length of the
contour $\Gamma$) and ${\det}'\hat {\cal N}$ is the regularized determinant of the 
Neumann jump operator with the zero mode removed. A simple heuristic derivation 
of the surgery formula is given in Appendix E.
This formula allows one to represent the final result for $F_2^{(q)}$ in the form
\beq\label{F2q}
F_{2}^{(q)}=\frac{1}{2}\log {\det }' \hat {\cal N}-\frac{1}{2} \log P +c_2.
\eeq
The constant $c_2$ is determined below from the comparison with the result for the circle.
The explicit expression of this quantity in terms the 
exterior and interior conformal maps are:
\beq\label{F2q1a}
\begin{array}{lll}
F_{2}^{(q)}&=&\displaystyle{\frac{1}{24\pi}\oint_{\Gamma}
\Bigl (\psi_{\rm ext}\p_n \psi_{\rm ext} -
\psi_{\rm int}\p_n \psi_{\rm int}+2\kappa (\psi_{\rm ext}-\psi_{\rm int})
\Bigr )ds\, + \mbox{const}}
\\ && \\
&=&\displaystyle{\frac{1}{24\pi} \oint_{|w|=1} \Bigl (
\phi _{{\rm int}}
\p_n \phi _{{\rm int}}
-\phi _{{\rm ext}}
\p_n \phi _{{\rm ext}} +
2 (\phi _{{\rm int}}\! -\! \phi _{{\rm ext}})
\Bigr )|dw| + \mbox{const}}.
\end{array}
\eeq

Let us point out yet another instructive representation of $F_2^{(q)}$. As is shown in
Appendix D, it is given by logarithm of the Fredholm determinant associated to the domain
$\DD$ (see \cite{Schiffer} for a comprehensive treatment of Fredholm eigenvalues of the
domain $\DD$). Using equation (\ref{double9}) from Appendix D, one can 
write (\ref{F2q}) in the form
\beq\label{F2q1}
F_2^{(q)}=-\frac{1}{2}\, \log {\det}(\hat {\cal I}+\hat {\cal V}) +\mbox{const},
\eeq
where $\hat {\cal I}$ is the identity operator and 
$\hat {\cal V}$ is the double layer potential operator
\beq\label{F2q2}
\hat {\cal V}f(z)=\frac{1}{\pi}\pvint \, f(\xi )\p_{n_{\xi}}
\log |\xi -z| |d\xi |, \quad z\in \Gamma 
\eeq
also known as the Neumann-Poincar\'e operator. 
Here $\displaystyle{\pvint}$ is the principal value integral. Note that the signs
in front of $\log \det$ in (\ref{F2q}) and (\ref{F2q1}) are opposite. 
The relation between the Fredholm determinant associated with a plane 
domain and determinants of the Laplace operators was mentioned in 
\cite{TT04} (Corollary 4.12).

\subsection{Examples and particular cases}

\paragraph{The case of no potential: $W=0$.} In this case, our formulas simplify:
\beq\label{no1}
\begin{array}{l}
\displaystyle{\rho_0=\frac{1}{2\pi}\, |w'_{\rm ext}|},
\\ \\
\displaystyle{\rho_1=\frac{\beta -1}{4\pi \beta} \, \hat {\cal N}\log |w'_{\rm ext}|},
\\ \\
\displaystyle{\rho_2=2\pi \left (\frac{\beta -1}{4\pi \beta}\right )^2 \!
\hat {\cal N} \frac{1}{|w'_{\rm ext}|}\, \hat {\cal N}\log |w'_{\rm ext}|}
\end{array}
\eeq
(this is obtained by iteration of equation (\ref{it4}) with $W=0$) and
\beq\label{no2}
\begin{array}{l}
F_0=\beta  \log r,
\\ \\
F_1 = -(\beta \! -\! 1) \log (2\pi r) +c_1,
\\ \\
\displaystyle{F_2^{(cl)}=\frac{(\beta -1)^2}{8\pi \beta}\oint_{\Gamma}
\log |w'_{\rm ext}| \, \hat {\cal N} \log |w'_{\rm ext}|\, ds +\mbox{const}}.
\end{array}
\eeq
The constant $c_1$ is given below in (\ref{c1}).
The quantum part of $F_2$ is given by (\ref{F2q}).
Note that if $W=0$, then contour-dependent contributions of 
$F_1$ and of the classical part of $F_2$ vanish at $\beta =1$.

\paragraph{$\Gamma$ is a circle and $W=0$.} 
In the case when 
$\Gamma$ is a circle of radius $r$ 
and $W=0$ the integral for the partition function had been explicitly 
evaluated \cite{Dyson}:
\beq\label{ZNa}
Z_N^{(0)} =\oint _{|z_1|=r} \! \ldots \! \oint _{|z|_N=r}
\prod_{i<j}|z_i -z_j |^{2\beta}
\prod_{k=1}^{N}|dz_k |=r^{\beta N^2 +(1-\beta )N} (2\pi )^N
\frac{\Gamma (1+N\beta )}{\Gamma^N(1+\beta )}.
\eeq
The conformal maps are $w_{\rm int}(z)=w_{\rm ext}(z)=z/r$. 
The density $\rho_0$ is constant: $\rho_0=1/(2\pi r)$.
The large $N$ expansion is of the form (\ref{large1}) with
\beq\label{ZNb}
\begin{array}{l}
F_0=\beta  \log r,
\\ \\
\displaystyle{
F_1=-(\beta -1) \log \frac{re}{\beta}+ \log \frac{2\pi}{\Gamma (\beta )},}
\\ \\
F_2 =\log \sqrt{\beta}.
\end{array}
\eeq
(The expansion is made under assumptions that $N\to \infty$ together with
$N\beta \to \infty$, so it is essential here that $\beta \neq 0$.)
Note that for the circle we have $\hat {\cal N}=2\hat {\cal N}^+$ (see (\ref{N1})) 
and $\det '\hat{\cal N}^+$
is known from \cite{EW91,GG07}: $\det '(2\hat 
{\cal N}^+) =\pi r$, so $F_2$ does not depend on $r$.
Comparison with (\ref{no2}) allows one to find the constants $c_1$, $c_2$:
\beq\label{c1}
c_1=(\beta -1) \log \frac{2\pi \beta}{e} +\log \frac{2\pi}{\Gamma (\beta )},
\eeq
\beq\label{c2}
c_2=\frac{1}{2}\, \log (2\beta ).
\eeq
A complete large $N$ expansion of (\ref{ZNa}) is readily available. 

\paragraph{The case $\beta =1$.} The value $\beta =1$ in (\ref{ZN}) is special. 
In this case the partition function has the determinant representation
\beq\label{ZNab}
Z_N^{(\beta =1)}=N! \det_{N\times N}M_{ij}, \qquad 1\leq i,j\leq N,
\eeq
where
$M_{ij}$ is the matrix of moments:
\beq\label{ZNc}
M_{ij}=\oint_{\Gamma} z^{i-1}\bar z^{j-1} e^{W(z)}|dz|.
\eeq
The proof is a simple calculation which 
is almost the same as for the 
circular Dyson ensemble. 

The determinant representation implies
the integrable structure. Let the potential $W$
be of the form
$$
W(z)=W^{(0)}(z)+\sum_{k\geq 1}(t_k z^k +\bar t_k \bar z^k),
$$
where $W^{(0)}$ is a fixed background potential and $t_k$ are varying parameters. 
It can be proved (see, e.g., \cite{Z04}) that the partition function $Z_N$ as a function
of the ``times'' $t_k, \bar t_k$ and $N$ is the tau-function of the 2D Toda lattice
hierarchy \cite{UT84}. As such, it satisfies the full set of bilinear Hirota equations.

\paragraph{The case $W(z)=(1-\beta )\log |w'_{\rm ext}(z)|$.}
The partition functions reads
\beq\label{ZN1}
Z_N =\oint _{\Gamma} \! \ldots \! \oint _{\Gamma}
\prod_{i<j}|z_i -z_j |^{2\beta}
\prod_{k=1}^{N}
|w'_{\rm ext}(z_k)|^{1-\beta}|dz_k |.
\eeq
In this case the potential depends on the contour and one should use the general
formula (\ref{it8}). One can see that
the contour-dependent contributions to
$F_1$ and to the classical part of $F_2$ vanish while the quantum part of 
$F_2$ is given by the same
formula (\ref{F2q1}). Therefore, the result for $Z_N$ as $N\to \infty$ is given entirely
in terms of the Fredholm determinant associated to the domain $\DD$:
\beq\label{ZN2}
Z_N =C r^{\beta N^2}({\det} (\hat {\cal I}+\hat {\cal V}))^{-1/2}(1+O(1/N)).
\eeq

\section{Correlation functions of vertex operators}

Along with the density and potential the operators 
$$
V_{\alpha}(z)=\prod_{j=1}^{N}(z-z_j)^{\alpha}\,,
\quad \quad
V_{\alpha}(z, \bar z)=\prod_{j=1}^{N}|z-z_j|^{2\alpha}
$$
are of interest. Following the analogy with conformal field
theory, we call them vertex operators. One can write 
$
V_{\alpha}(z, \bar z)=e^{-\frac{\alpha}{\beta \hbar}\varphi (z)}
$,
$
V_{\alpha}(z)=e^{-\frac{\alpha}{\beta \hbar}\phi (z)}
$,
where $\phi (z)$ is the holomorphic part of the potential
$\varphi$:
\beq\label{cfvo1}
\phi (z)=-\beta \hbar \sum_j \log (z-z_j)\,,
\quad \quad
\varphi (z)=\phi (z)+\overline{\phi (z)}.
\eeq
Below in this section, we find correlation functions 
of these fields in the leading order setting the 
external potential $W$ to be zero and choosing $z$ in the exterior domain.

\paragraph{Correlation functions of $V_{\alpha}(z)$.}
Using the well known formula for the mean value of 
exponential function, we can write, in the large $N$ limit:
\beq\label{V1}
\lbracket V_{\alpha}(z)\rbracket =
\exp \left (-\frac{\alpha}{\beta \hbar}\lbracket \phi (z)\rbracket
+\frac{\alpha^2}{2\beta^2 \hbar^2}\lbracket \phi ^2 (z)
\rbracket_{{\rm c}}+O(\hbar ) \right ).
\eeq
The correlation functions of the field
$\phi$ can be easily found from those of the field $\varphi$
by extracting the holomorphic part. 
Now we set the point $z$ in the exterior of $\Gamma$.
In this case
$$
\lbracket \phi (z)\rbracket =-\beta 
\log (rw_{{\rm ext}}(z))+\frac{\hbar}{2}(\beta -1)
\log (rw'_{{\rm ext}}(z))+O(\hbar^2)
$$
(the second term comes from (\ref{next1}))
and
$$
\frac{\lbracket \phi (z)\phi (\zeta )\rbracket_{{\rm c}}}{\beta \hbar^2}
\, \cong \, 
\log \frac{rw_{{\rm ext}}(z)\! -\! rw_{{\rm ext}}(\zeta )}{z-\zeta}.
$$
Clearly, the r.h.s. becomes $\log (rw'_{{\rm ext}}(z))$ as
$\zeta \to z$.
Therefore, one obtains from (\ref{V1}):
\beq\label{V2}
\lbracket V_{\alpha}(z)\rbracket  \, \cong \,
(rw_{{\rm ext}}(z))^{\alpha N}
(rw'_{{\rm ext}}(z))^{\frac{\alpha}{2\beta}(\alpha +1-\beta )}.
\eeq

Correlation functions of multiple products of the fields
$V_{\alpha}$ can be obtained in a similar way. For example,
\beq\label{V3}
\lbracket V_{\alpha_1}(z_1)V_{\alpha_2}(z_2)\rbracket  
\, \cong \,
\lbracket V_{\alpha_1}(z_1)\rbracket \lbracket 
V_{\alpha_2}(z_2)\rbracket 
\left (\frac{rw_{{\rm ext}}(z_1)\! -\! 
rw_{{\rm ext}}(z_2)}{z_1 -z_2}\right )^{\frac{\alpha_1 \alpha_2}{\beta}}
\eeq
and, more generally,
\beq\label{V4}
\lbracket \prod_{p=1}^{m}V_{\alpha_p}(z_p)\rbracket 
=\prod_{p=1}^{m}\lbracket V_{\alpha_p}(z_p)\rbracket
\prod_{p<q}^{m}
\left (\frac{rw_{{\rm ext}}(z_p)\! -\! 
rw_{{\rm ext}}(z_q)}{z_p -z_q}\right )^{\frac{\alpha_p \alpha_q}{\beta}}.
\eeq

Formulas for correlation functions of these fields for $z$
in the interior of $\Gamma$ 
include not only the interior conformal map $w_{\rm int}(z)$
but also the holomorphic part of the harmonic continuation
of the function $\log |w_{\rm out}'(z)|$ to the interior.
We do not present them here. 

\paragraph{Correlation functions of $V_{\alpha}(z, \bar z)$.}
Similarly to (\ref{V1}), we write
\beq\label{V1a}
\lbracket V_{\alpha}(z, \bar z)\rbracket =
\exp \left (-\frac{\alpha}{\beta \hbar}\lbracket \varphi (z)\rbracket
+\frac{\alpha^2}{2\beta^2 \hbar^2}\lbracket \varphi ^2 (z)
\rbracket_{{\rm c}}+O(\hbar ) \right )
\eeq
and make use of (\ref{next1}) and (\ref{lead6}). 
Plugging
$$
\lbracket \varphi (z)\rbracket =-2\beta 
\log |rw_{{\rm ext}}(z)|+ \hbar (\beta -1)
\log |rw'_{{\rm ext}}(z)|+O(\hbar^2),
$$
$$
\lbracket \varphi ^2 (z)\rbracket _{\rm c}=2\beta \hbar^2
\log \left | \frac{rw_{\rm ext}'(z)}{1-
|w_{\rm ext}(z)|^{-2}} \right |
$$
into (\ref{V1a}), we obtain:
\beq\label{V2a}
\lbracket V_{\alpha}(z, \bar z)\rbracket  \, \cong \,
|rw_{{\rm ext}}(z)|^{2\alpha N}
|rw'_{{\rm ext}}(z)|^{\frac{\alpha}{\beta}(\alpha +1-\beta )}
\left (1-|w_{\rm ext}(z)^{-2}\right )^{-\frac{\alpha^2}{\beta}}
\eeq
and
\beq\label{V4a}
\lbracket \prod_{p=1}^{m}V_{\alpha_p}(z_p, \bar z_p)\rbracket 
=\prod_{p=1}^{m}\lbracket V_{\alpha_p}(z_p, \bar z_p)\rbracket
\prod_{p<q}^{m}
\left |\frac{rw_{{\rm ext}}(z_p)\! -\! 
rw_{{\rm ext}}(z_q)}{(z_p -z_q)(1-w_{\rm ext}^{-1}(z_p)
\overline{w_{\rm ext}^{-1}(z_q)}}
\right |^{\frac{2\alpha_p \alpha_q}{\beta}}.
\eeq

These formulas match the correlation functions of vertex operators 
in the boundary conformal field theory with the central charge
$c=1-6\Bigl (\beta^{1/2}-\beta^{-1/2}\Bigr )^2$.

\section{Concluding remarks}

We have introduced and studied the statistical 
model of logarithmic gas (the $\beta$-ensemble) on a general smooth closed contour 
$\Gamma \subset \CC$ 
of arbitrary form. This model generalizes Dyson's circular $\beta$-ensembles. We have developed
a systematic procedure of the large $N$ expansion based on an identity for correlation
functions often called loop equation. Its origin  
is reparameterization invariance of the
contour integrals representing the partition function. 
A combination of the loop equation with dilatation transformations of the contour allows us
to compute first
three contributions to the free energy which do not vanish
as $N \to \infty$, in orders $O(N^2)$, $O(N)$ and $O(1)$ respectively.
They are found explicitly and expressed through geometric
characteristics of the contour and the spectral geometry
of the interior and exterior domains. In particular, 
the most interesting
contribution of the order of $O(1)$ is expressed
in terms of the Neumann jump operator $\hat {\cal N}$ 
which takes a function on $\Gamma$ to the 
difference of normal derivatives of its harmonic continuations to the interior and exterior
of the contour. 
The form of the final results is in parallel to that
of 2D logarithmic gases where the $O(1)$ part of the free energy is expressed through the
spectral determinant of the Laplace operator of the supporting domain 
\cite{WZ03a,WZ06}. 

We have also found correlation functions 
of the charge density and electrostatic 
potential in the leading order in $N$. They are expressed through Green's
functions of the Laplace operators in the interior and exterior domains. 

At last, we comment on the relation of our results to 
2D conformal field theory (CFT). 
First, we would like to emphasize that the holomorphic extension of the
stress tensor $T(z)$ is similar to that of CFT
and the important fact that the loop equation 
turns out to be equivalent 
to a boundary condition for the holomorphic extension 
of the stress tensor of the logarithmic gas.
The loop equation states that 
the tangential-normal component of the stress 
is continuous across the contour. 
This condition is to be compared to the conformal boundary condition for the stress
tensor in the boundary CFT. In that case, the stress is defined
only from one side of the boundary where 
its tangential-normal component vanishes. Besides, in CFT the 
stress tensor is traceless everywhere including the boundary.
With the conformal boundary condition, under a change of the geometry 
(a deformation of the contour) correlation 
functions in CFT transform covariantly under
conformal transformations of the domain bounded by the contour.
This is not the case in our model,
and so the similarity with CFT is not complete.  
In our case, the stress tensor $T$ has a non-vanishing trace located exclusively
on the boundary.
As a result, the variation of the free energy under small deformations of
the contour is expressed through the normal-normal component of $T$. Also,
it is important that the stress 
is well-defined not only in the large $N$ limit but also 
for any finite $N$, so one may regard it
as a natural ``discretization'' of the
stress tensor of field theory when we treat the gas as a continuous media. 

Like in the CFT, the $O(1)$ part of the free energy, $F_2$, 
is expressed through the free
energies of 2D Bose field. However, contrary to the CFT, in our case, the support of the Bose
field are two domains, the exterior and the interior of the contour. 
They are expressed through the regularized determinants of Laplace operators 
in the interior and exterior domains (equation (\ref{q3a}))
despite that, the physical matter is confined to the contour.

Finally, we mention yet another interpretation of $F_2$.
It is the free energy of
1D fermions $\psi$
confined on the contour with the action $\oint_{\Gamma} \psi \hat {\cal N}\psi ds$.

Despite of the differences,
the correlation functions of ``vertex operators'' 
(exponential functions of the electrostatic potential created by the charges) 
located in the exterior domain 
turn out to be similar to the correlators of primary fields in CFT.

When this work was completed, we became aware of the recent paper
\cite{Johansson21}, where similar results for $\beta =1$ were obtained 
by more rigorous mathematical methods.

\section*{Appendix A: Normal and tangential derivatives}
\addcontentsline{toc}{section}{\hspace{6mm}Appendix A: Normal and tangential derivatives}
\def\theequation{A\arabic{equation}}
\def\theHequation{\theequation}
\setcounter{equation}{0}

Given a smooth closed curve $\Gamma$, let $\tau (z)=dz/|dz|$ be the unit
tangent vector at $z\in \Gamma$ represented as
a complex number pointing in positive (counterclockwise)
direction. Then $\nu (z)=-i\tau (z)$ is the outward
pointing unit normal vector to the curve $\Gamma$ at the point $z$.
Clearly, $|\nu(z)|=1$ for $z\in \Gamma$, i.e., $\overline{\nu (z)}=\nu ^{-1}(z)$.
In terms of the conformal map $w(z)=w_{\rm ext}(z)$ from the exterior of $\Gamma$ to the
exterior of the unit circle (we normalize it by the conditions $w(\infty )=\infty$ and 
$w'(\infty )$ is real positive) we have
\beq\label{nu}
\nu (z)=|w'(z)|\frac{w(z)}{w'(z)}.
\eeq
The line element along $\Gamma$ is
$$
ds = |dz|=\sqrt{dz d\bar z}
=\frac{dz}{i\nu (z)}=i\nu (z) d\bar z.
$$
Given any function $f(z)=f(z, \bar z)$ in the
complex plane, its normal and tangential derivatives
on $\Gamma$ are:
\beq\label{ap0}
\begin{array}{l}
\p_n f (z)=\nu (z) \p_z f(z) +\overline{\nu (z)} \p_{\bar z}f(z),
\\ \\
\p_s f (z)=i\nu (z) \p_z f(z) -i\overline{\nu (z)} \p_{\bar z}f(z).
\end{array}
\eeq

We write $\nu (z)=e^{i\theta}$, where $\theta$ is the angle between the
normal vector and the real axis.
The curvature is given by
\beq\label{curvature}
\kappa = \frac{d\theta}{ds}=i\nu \,\p_s \bar \nu .
\eeq
The equivalent formula
\beq\label{kappa}
\kappa (z)=\p_n \log \left |\frac{w(z)}{w'(z)}\right |
\eeq
is an immediate consequence of the definition.

The Laplace operator $\Delta = \p_x^2+\p_y^2=4\p_z \p_{\bar z}$
at a point $z\in \Gamma$ is
expressed in terms normal and tangential derivatives as
$$
\Delta = \p_{n}^{2}+\p_{s}^{2}+\kappa \p_n.
$$

Let $f,g$ be any two functions in the plane defined in a strip-like neighborhood 
of the contour $\Gamma$.
The following formulas for $z\in \Gamma$ are often used
in the calculations:
\beq\label{ap1}
\displaystyle{
2{\cal R}e \left ( \nu ^2 \p f \, \p g\right )=
\frac{1}{2}( \p_n f \p_n g - \p_s f \p_s g)},
\eeq
\beq\label{ap2}
\displaystyle{
2{\cal I}m \left ( \nu ^2 \p f \, \p g\right )=
-\, \frac{1}{2}( \p_n f \p_s g + \p_s f \p_n g)}.
\eeq
Assuming that $\Delta f =0$, we also have:
\beq\label{ap3}
\displaystyle{
2{\cal R}e \left ( \nu ^2 \p^2 f \right )=
-\p_{s}^{2}f - \kappa \p_n f},
\eeq
\beq\label{ap4}
\displaystyle{
2{\cal I}m \left ( \nu ^2 \p^2 f \right )=
-\p_s \p_n f + \kappa \p_s f}.
\eeq

For a function $f=u+iv$ analytic in a vicinity of the contour the Cauchy-Riemann identities
read $\p_n u=\p_s v$, $\p_n v=-\p_s u$. 

\section*{Appendix B: The Green's functions}

\addcontentsline{toc}{section}{\hspace{6mm}Appendix B: The Green's functions}
\def\theequation{B\arabic{equation}}
\def\theHequation{\theequation}
\setcounter{equation}{0}

The Green's function of the Dirichlet boundary value
problem in a domain $\DD \subset \CC$ is a solution to the equation
$\Delta _z G(z, \zeta )=\Delta_{\zeta}(z,\zeta )G(z, \zeta )=2\pi
\delta^{(2)}(z-\zeta )$ vanishing on $\p \DD$. The
Green's function is symmetric: $G(z, \zeta )=G(\zeta , z)$.
As $\zeta \to z$, it has the logarithmic singularity
$G(z, \zeta )=\log |z-\zeta |+\ldots $. If $\DD$ is simply
connected, then the Green's function can be expressed through
the conformal map $w(z)$ from $\DD$ onto the unit disk by
the formula
$$
G(z, \zeta ) =\log \left |
\frac{w(z) - w(\zeta )}{1-w(z)\overline{w(\zeta )}}
\right |.
$$

Let $\Gamma$ be a closed contour in the plane
encircling the origin.
Given such a contour,
let $\DD$ be the interior
domain and $G_{{\rm int}}$, $G_{{\rm ext}}$ be the Green's
functions in $\DD$ and $\Dc$ respectively. We also introduce
the conformal maps $w_{{\rm int}}(z)$ from $\DD$
onto the unit disk and the conformal map $w_{{\rm ext}}(z)$
from $\Dc$ onto the exterior of the unit disk such that
$w_{{\rm int}}(0)=0$, $w_{{\rm ext}}(\infty )=\infty$ and
their derivatives at $0$ and $\infty$ are real positive.

The formulas
\beq\label{Dirichlet}
\begin{array}{l}
\displaystyle{
f_H (z) \, =\, \frac{1}{2\pi} \oint_{\Gamma}f(\xi )
\p_{n}^{+} G_{{\rm int}} (z, \xi )
|d\xi |,}
\\ \\
\displaystyle{
f^H (z) =-\, \frac{1}{2\pi} \oint_{\Gamma}f(\xi )
\p_{n}^{-} G_{{\rm ext}} (z, \xi )
|d\xi |}
\end{array}
\eeq
provide solutions to the Dirichlet boundary value
problems in $\DD$ and $\Dc$, i.e., they give harmonic continuations of a function $f$
from the contour to its interior and exterior
respectively. Here
$\p_{n}^{\pm}$ mean normal derivatives taken from inside
(outside), with the normal vector looking outside in both cases.
The Poisson kernel
$\p_{n_\xi} G (z, \xi )$ (here $G$ is $G_{{\rm int}}$ or $G_{{\rm ext}}$ and
$\xi \in \Gamma =\p \DD$)
is expressed through the conformal map $w(z)$ (correspondingly, $w_{{\rm int}}$
or $w_{{\rm ext}}$) as follows:
$$
\begin{array}{lll}
\p_{n_{\xi}}G(z, \xi )&=&
\displaystyle{\frac{|w'(\xi )|\, (1-|w(z)|^2)}{|w(\xi )-w(z)|^2}\, =\, 
|w'(\xi )|{\cal R}e \left (
\frac{w(\xi )+w(z)}{w(\xi )-w(z)}\right )}
\\ && \\
&=&\displaystyle{
|w'(\xi )| \left ( \frac{w(z)}{w(\xi )-w(z)} -
\frac{( \overline{w(z)} )^{-1}}{w(\xi )-(\overline{w(z)} )^{-1}}\right )}.
\end{array}
$$

The variation of the Green's function under small deformation of the contour is given by the
Hadamard formula
\beq\label{Hadamard}
\delta G(a,b)=\mp \frac{1}{2\pi}\oint_{\Gamma}\p_n G(a, \xi )\p_n G(b, \xi )\delta n (\xi )
|d\xi |,
\eeq
where $\delta n (\xi )$ is the normal displacement of the boundary along the normal vector
at the point $\xi \in \Gamma$ and minus (plus) sign corresponds to the interior (exterior) problem.

\section*{Appendix C: The Neumann jump operator}
\addcontentsline{toc}{section}{\hspace{6mm}Appendix C: The Neumann jump operator}
\def\theequation{C\arabic{equation}}
\def\theHequation{\theequation}
\setcounter{equation}{0}

Given a smooth closed contour $\Gamma$ in the plane, one defines the Neumann
jump operator as
$$
\hat {\cal N} f(z) = \p_{n}^{+} f_H (z) -\p_{n}^{-} f^H (z)\,, \quad z\in \Gamma .
$$
Since
$$
\oint_{\Gamma} f\p_n^{+}g_H ds =\oint_{\Gamma} g\p_n^{+}f_H ds ,
\qquad
\oint_{\Gamma} f\p_n^{-}g^H ds =\oint_{\Gamma} g\p_n^{-}f^H ds
$$
by virtue of Green's theorem, 
the operator $\hat {\cal N}$ is symmetric (self-adjoint) with respect to the 
scalar product   
$\oint_{\Gamma} fg ds$
of functions $f$, $g$ on $\Gamma$.

The Neumann jump operator has a zero mode: a constant function. In the space of
functions $f$ such that $\oint_{\Gamma}fds=0$ the operator $\hat {\cal N}$
is non-degenerate. 
The kernel of the inverse operator in this space is
$-\frac{1}{2\pi}\log |z-\xi |$. Indeed, assuming that $\oint_{\Gamma}fds=0$,
let us define $g(z)$ by
\beq\label{ap5}
g (z)=-\, \frac{1}{2\pi}\oint_{\Gamma} \log |z-\xi |f(\xi )\, |d\xi |\,,
\quad z\in \Gamma ,
\eeq
then $\hat {\cal N} g (z)=f(z)$ by the property of potentials of a simple layer.

More precisely, let $\hat {\cal K}$ be the operator defined by
\beq\label{ap6}
\hat {\cal K}f(z)=-\frac{1}{2\pi}\oint_{\Gamma}\log |z-\xi |f(\xi )\, |d\xi |
\eeq
on the space of all functions on $\Gamma$ (the simple layer operator). 
It is not difficult to see that the following relations hold true:
\beq\label{ap8}
\hat {\cal N}\hat {\cal K}f=f -\frac{1}{2\pi}|w'|\oint_{\Gamma}fds,
\eeq
\beq\label{ap9}
\hat {\cal K}\hat {\cal N}f=f -\frac{1}{2\pi}\oint_{\Gamma}|w'|f \, ds,
\eeq
where $w=w_{\rm ext}$, and
\beq\label{ap10}
\hat {\cal K}|w'|=\log (1/r).
\eeq
The first one easily follows from the properties of the simple layer potential. 
Let us outline the proof of the second one. The left hand side of it equals
$$
-\frac{1}{2\pi}\oint_{\Gamma} \log |\xi -z|\p_n^+ f_H (\xi )|d\xi |+
\frac{1}{2\pi}\oint_{\Gamma} \log |\xi -z|\p_n^- f^H (\xi )|d\xi |.
$$
Each integral here is a simple layer potential. Using the fact that the simple layer potential
is continuous across the contour, we can think that in the first integral $z$ approaches the
contour from the exterior ($z_-$) while in the second one $z$ approaches the contour
from the interior ($z_+$). Then the harmonic continuation of $\log |\xi -z_-|$ to the interior
is the function $\log |\xi -z_-|$ itself while the 
harmonic continuation of $\log |\xi -z_+|$ to the exterior is
$\log |\xi -z_+|-\log |w(\xi )|$. Therefore, using Green's theorem, we can transform
the left hand side of (\ref{ap9}) as follows:
$$
-\frac{1}{2\pi}\oint_{\Gamma} \log |\xi -z_-|\p_n^+ f_H (\xi )|d\xi |+
\frac{1}{2\pi}\oint_{\Gamma} \log |\xi -z_+|\p_n^- f^H (\xi )|d\xi |
$$
$$
=-\frac{1}{2\pi}\oint_{\Gamma} f(\xi )\p_{n_{\xi}}\log |\xi -z_-|\, |d\xi |+
\frac{1}{2\pi}\oint_{\Gamma} f(\xi )\p_{n_{\xi}}\log |\xi -z_+|\, |d\xi |
$$
$$\hspace{-2cm}
-\frac{1}{2\pi}\oint_{\Gamma}f(\xi )\p_{n_{\xi}}\log |w(\xi )|\, |d\xi |.
$$
The first two integrals on the right hand side are double-layer potentials. 
The properties of the double layer potentials are outlined in Appendix D.
The main property is that the double layer potential is discontinuous across
the contour (see (\ref{double1})). Equation (\ref{double1}) states that the difference of
the first two integrals in the right hand side above (the jump of the
double layer potential across the contour) 
is equal to $f(z)$. Finally, the second equation in (\ref{Dirichlet}) states that
the last integral provides the harmonic continuation of the function $f$ to infinity
and thus equals $f^H(\infty )$. Therefore, we see that the left hand side of 
(\ref{ap9}) is $f(z)-f^H(\infty )$ which is the same as the right hand side.

In the case when $\Gamma$ is a circle
$$
\p_{n}^{+}f_H (z)=- \, \p_{n}^{-}f^H (z),
$$
and
\beq\label{N1}
\hat {\cal N}f =2\p_{n}^{+}f_H =2\hat {\cal N}^+ f.
\eeq

\section*{Appendix D: Double layer potentials and Fredholm eigenvalues of planar domains}

\addcontentsline{toc}{section}{\hspace{6mm}Appendix D: Double layer potentials
and Fredholm eigenvalues of planar domains}
\def\theequation{D\arabic{equation}}
\def\theHequation{\theequation}
\setcounter{equation}{0}

\paragraph{Double layer potentials.}
Given a function $f$ on $\Gamma$, one can construct
the double layer potential
defined by
$$
V_f (z)=\frac{1}{\pi}\oint_{\Gamma}f(\xi )\p_{n_{\xi}}\log |\xi -z|
|d\xi |.
$$
It is harmonic everywhere in the plane except for the contour $\Gamma$
where it has a jump.
Its behavior across the boundary is given by
\beq\label{double1}
V_f (z) =\left \{
\begin{array}{l}
\displaystyle{
\,\,\, f(z)+\frac{1}{\pi}\pvint \,
f(\xi )\p_{n_{\xi}}\log |\xi -z| |d\xi |\,, \quad z\to z_+}
\\ \\
\displaystyle{
\phantom{aaaaaaa} \frac{1}{\pi}\pvint \,
f(\xi )\p_{n_{\xi}}\log |\xi -z| |d\xi |\,, \quad z\in \Gamma}
\\ \\
\displaystyle{
-f(z)+\frac{1}{\pi}\pvint\,
f(\xi )\p_{n_{\xi}}\log |\xi -z| |d\xi |\,, \quad z\to z_-},
\\ \\
\end{array} \right.
\eeq
where $z \to z_{\pm}$ means that one approaches the point
$z\in \Gamma$ from inside and outside respectively and
$\displaystyle{ \pvint}$ is the principal
value integral. 
Using the Green's theorem, we can represent (\ref{double1}) as simple layer potentials:
\beq\label{double3}
\begin{array}{l}
\displaystyle{
V_f(z_+)=\frac{1}{\pi}\oint_{\Gamma} 
\log |\xi -z| \p_{n}^{-}f^H (\xi ) |d\xi | +2f^H (\infty )},
\\ \\
\displaystyle{
V_f(z_-)=\frac{1}{\pi}\oint_{\Gamma} \log |\xi -z| \p_{n}^{+}f_H (\xi ) |d\xi |}.
\end{array}
\eeq
Let us introduce the operator $\hat {\cal V}$ acting in the space
of functions on $\Gamma$ in the following way:
\beq\label{double2}
\hat {\cal V}f(z)=\frac{1}{\pi}\pvint \, f(\xi )\p_{n_{\xi}}
\log |\xi -z| |d\xi |, \quad z\in \Gamma .
\eeq
Note that it is not self-adjoint.
In this notation
$$V_f(z_+)=(\hat {\cal I}+\hat {\cal V})f(z), 
\quad V_f(z_-)=-(\hat {\cal I}-\hat {\cal V})f(z).
$$

Applying the Neumann jump operator $\hat {\cal N}$ to both sides of 
(\ref{double3}) and using
the properties of the simple layer potentials, we have:
\beq\label{double4}
\hat {\cal N}(\hat {\cal I}+\hat {\cal V})f = -2\p_{n}^- f^H,
\quad
\hat {\cal N}(\hat {\cal I}-\hat {\cal V})f = 2\p_{n}^+ f_H
\eeq
or, in the operator form,
\beq\label{double5}
\hat {\cal N}(\hat {\cal I}+\hat {\cal V}) = -2\hat {\cal N}^-,
\quad
\hat {\cal N}(\hat {\cal I}-\hat {\cal V}) = 2\hat {\cal N}^+.
\eeq

For completeness, we present here some other properties of the simple and double
layer potentials (see \cite{Taylor}).
The adjoint operator $\hat {\cal V}^{\dag}$
acts as
\beq\label{double2a}
\hat {\cal V}^{\dag}f(z)=\frac{1}{\pi}\pvint \, f(\xi )\p_{n_{z}}
\log |\xi -z| |d\xi |, \quad z\in \Gamma .
\eeq
Equations (\ref{double5}) imply the following values of normal
derivatives of simple layer potential:
\beq\label{double5b}
\begin{array}{l}
-2\p_n^+ \hat {\cal K}f(z)=-f(z)+\hat {\cal V}^{\dag}f(z),
\\ \\
-2\p_n^- \hat {\cal K}f(z)=f(z)+\hat {\cal V}^{\dag}f(z).
\end{array}
\eeq
Summing the first relation in (\ref{double5}) with the adjoint second one, we get
\beq\label{double5a}
\hat {\cal N}\hat {\cal V}=\hat {\cal V}^{\dag}\hat {\cal N}.
\eeq
Taking the difference of the relations (\ref{double5}), we obtain the operator relation
$$
\hat {\cal N}^+(\hat {\cal I}+\hat {\cal V})=-\hat {\cal N}^-(\hat {\cal I}-\hat {\cal V}),
$$
which is equivalent to 
\beq\label{double1a}
\p_{n}^{+}V_f(z)=\p_{n}^{-}V_f(z)
\eeq
(the normal derivative of the double layer potential is continuous across the boundary).
With the help of (\ref{double3}), the continuity relation (\ref{double1a}) can be 
also represented 
in the operator form as follows:
\beq\label{double1b}
\hat {\cal N}^+ \hat {\cal K}\hat {\cal N}^- =
\hat {\cal N}^- \hat {\cal K}\hat {\cal N}^+.
\eeq
The proof of this operator relation is based on (\ref{ap8}), (\ref{ap9}).

\paragraph{Fredholm eigenvalues of the domain $\DD$.}
Eigenvalues $\lambda_{\alpha}$ of the operator $\hat {\cal V}$ are called
Fredholm eigenvalues of the domain $\DD$ \cite{Schiffer}.
The eigenvalue equation can be written in the form
\beq\label{double6}
\frac{1}{\pi}
\pvint \, \phi_{\alpha}(\xi )
\, \p_{n_{\xi}} \log |\xi -z| |d\xi |=\lambda_{\alpha}\phi_{\alpha}(z).
\eeq
It is easy to see that the constant function $\phi_0=1$ is the eigenfunction with the
eigenvalue $\lambda_0=1$. For $\alpha \neq 0$, one can show \cite{Schiffer} 
that if $\lambda_{\alpha}$ is an eigenvalue, then
$-\lambda_{\alpha}$ is also an eigenvalue and $|\lambda_{\alpha}|\leq 1$. 
Moreover, estimates for the eigenvalues \cite{MS17,AKMP20,KPS07} imply that 
for smooth enough contours 
$\displaystyle{\sum_{\alpha}|\lambda_{\alpha}|<\infty }$.
The Fredholm determinant associated to the domain $\DD$,
or rather to the contour $\Gamma$, is
\beq\label{double7}
\det (\hat {\cal I}+\hat {\cal V}) 
= \prod_{\alpha}(1+\lambda_{\alpha}).
\eeq
For smooth enough contours, the infinite product is convergent and no regularization 
is required (see \cite{MS17,AKMP20}). We also have
\beq\label{double7a}
\det (\hat {\cal I}+\hat {\cal V})=2{\det}'(\hat {\cal I}-\hat {\cal V}),
\eeq
where prime in the right hand side means, as usual, that the zero eigenvalue is removed. 

\paragraph{Determinant of the Neumann jump operator and Fredholm determinant.}
Equations (\ref{double5}) imply that
\beq\label{double8}
{\det}' \hat {\cal N}\, {\det} (\hat {\cal I}+\hat {\cal V})=2{\det}'(-2\hat {\cal N}^-),
\eeq
or
\beq\label{double8a}
{\det}' \hat {\cal N}\, {\det}' (\hat {\cal I}-\hat {\cal V})={\det}'(2\hat {\cal N}^+),
\eeq
where the modified determinants ${\det}' \hat {\cal N}$, ${\det}' (2\hat {\cal N}^{\pm})$ are 
regularized by means of the $\zeta$-functions (see Appendix F).
As is proved in \cite{EW91},  ${\det}'(2\hat {\cal N}^+)=-{\det}' (-2\hat {\cal N}^{-})=
\frac{1}{2}\, P$, where $P=\oint_{\Gamma}ds$ is length of the contour
$\Gamma$. Therefore, the regularized determinant of the Neumann jump operator is expressed
through the Fredholm determinant of the domain $\DD$ as follows:
\beq\label{double9}
\log {\det}' \hat {\cal N}=-\log {\det} (\hat {\cal I}+\hat {\cal V})+\log P.
\eeq

The formulas (\ref{double8}), (\ref{double8a}) require some justification. We regularize the 
determinant of the left hand side of (\ref{double5}) as
$$
\lim_{\mu \to 0}\det \Bigl ((\hat {\cal N}_{\mu}+\varepsilon \hat {\cal P}_0)
(\hat {\cal I}+\hat {\cal V})\Bigr ),
$$
where $\hat {\cal N}_{\mu}=\hat {\cal N}+\mu \hat {\cal I}$ and $\hat {\cal P}_0$ is the
rank one projector on constant functions acting as
$$
\hat {\cal P}_0 f=\frac{1}{P}\oint_{\Gamma}fds.
$$
Since $\hat {\cal P}_0$ is a rank one operator, we have
$$
\det \Bigl ((\hat {\cal N}_{\mu}+\varepsilon \hat {\cal P}_0)
(\hat {\cal I}+\hat {\cal V})\Bigr )=\det \Bigl (\hat {\cal N}_{\mu}
(\hat {\cal I}+\hat {\cal V})\Bigr )\Bigl (1+\varepsilon \mbox{tr}\, (\hat {\cal N}_{\mu}^{-1}
\hat {\cal P}_0)\Bigr ).
$$
In our case, the operators $\hat {\cal N}_{\mu}$ and $\hat {\cal I}+\hat {\cal V}$
do not have a multiplicative anomaly \cite{KV95,Fried}, i.e. 
$\det \Bigl (\hat {\cal N}_{\mu}
(\hat {\cal I}+\hat {\cal V})\Bigr )=\det \hat {\cal N}_{\mu}
\det (\hat {\cal I}+\hat {\cal V})$. Therefore,
$$
\lim_{\mu \to 0}\det \Bigl ((\hat {\cal N}_{\mu}+\varepsilon \hat {\cal P}_0)
(\hat {\cal I}+\hat {\cal V})\Bigr )=\varepsilon \, {\det}' \hat {\cal N}\,
\det (\hat {\cal I}+\hat {\cal V}).
$$
We can write the first equation in (\ref{double5}) in the form
\beq\label{double5c}
(\hat {\cal N}_{\mu}+\varepsilon \hat {\cal P}_0 )(\hat {\cal I}+\hat {\cal V})=
-2\hat {\cal N}^- +\mu (\hat {\cal I}+\hat {\cal V})+\varepsilon \hat {\cal P}_0
(\hat {\cal I}+\hat {\cal V}).
\eeq
The determinant of the right hand side can be calculated as above
with the result
$$
\lim_{\mu \to 0}
\det \Bigl (-2\hat {\cal N}^- +\mu (\hat {\cal I}+\hat {\cal V})+\varepsilon \hat {\cal P}_0
(\hat {\cal I}+\hat {\cal V})\Bigr )=2\varepsilon \, {\det}' (-2\hat {\cal N}^-).
$$
This yields (\ref{double8}).

\section*{Appendix E: Variation of contour integrals}
\addcontentsline{toc}{section}{\hspace{6mm}Appendix E: Variations of contour integrals}
\def\theequation{E\arabic{equation}}
\def\theHequation{\theequation}
\setcounter{equation}{0}

Here we present the rules 
of variation of contour integrals over the simple closed curve $\Gamma$
under small deformations
of the curve. The variation of the curve is described by
the normal displacement $\delta n(z)$.

Consider a contour integral of a general form
$
\oint_{\Gamma} F(f, \p_n f)ds
$,
where $F$ is a fixed
function and the function $f$ may depend on the contour. Calculating the linear response
to the deformation of the contour,
one should vary all items in the integral independently
and add the results. There are four things to be varied:
the symbol $\oint_{\Gamma}$, the normal derivative $\p_n$, the line element $ds$
and the function $f$. By variation of $\oint_{\Gamma}$ we mean
integration of the old function over the new contour,
which gives the contribution
$$
\oint \delta n \, \p_n F \,ds.
$$
The change of the slope of the normal vector results in
$$
\delta \frac{\p}{\p n} =-\p_s (\delta n) \frac{\p}{\p s}.
$$
The rescaling of the line element gives
$$
\delta ds =\kappa \,\delta n ds.
$$
The variation of the function $f$ is case-dependent.
The procedure is clarified by a few examples which are used in the main text.

\paragraph{Example 1.}
\beq\label{examp1}
\delta \oint_{\Gamma} f(z)ds =
\oint_{\Gamma} \delta n (z) \Bigl (\p_n +\kappa (z)\Bigr )f(z)\, ds.
\eeq
The result holds provided the function $f$ itself does not change.
If it does, one should add the term $\oint_{\Gamma}
\delta f(z) ds$.

\paragraph{Example 2.}
$$
\begin{array}{lll}
\displaystyle{\delta \oint_{\Gamma} f(z)\p_n g(z) ds} &=&\displaystyle{
\oint_{\Gamma} \delta n(z) \,\p_n \Bigl (f(z) \p_n g(z)\Bigr )\, ds} 
\\&&\\
&&+\,\, \displaystyle{
\oint_{\Gamma} \delta n(z) \, f(z) \p_n g(z) \kappa (z)\, ds} 
\\&&\\
&&-\,\, \displaystyle{
\oint_{\Gamma} \p_s (\delta n(z)) f(z) \p_s g(z)\, ds},
\end{array}
$$
where the functions $f,g$ are fixed.
Taking the last integral by parts and using
the formula for the Laplace operator evaluated on the contour
$
\p_{n}^{2}+\p_{s}^{2} +\kappa \p_n =\Delta
$, 
we can represent the result
in the invariant form:
\beq\label{examp2}
\delta \oint_{\Gamma} f(z)\p_n g(z) ds =
\oint_{\Gamma} \delta n(z) \Bigl ((\nabla f , \nabla g )+
 f \Delta g \Bigr ) ds.
\eeq
Here $(\nabla f , \nabla g )=\p_n f \p_n g + \p_s f \p_s g$.
The same result can be obtained in an easier way by
reducing the contour integral to a 2D integral
by means of the Green theorem.

\paragraph{Example 3.}
Let us vary the integrals
$
\oint_{\Gamma} f(z)\p_n^+ g_H (z) ds$,
$\oint_{\Gamma} f(z)\p_n^- g^H (z) ds$, where $f,g$ are assumed to be 
contour-independent functions. However, 
the functions $g_H$, $g^H$ change when we change
the contour. We have:
\beq\label{examp6}
\p_n^+ \delta g_H = \p_n^+ \Bigl ( \p_n^+ (g\!-\! g_H )\, \delta n \Bigr )_H, \quad
\p_n^- \delta g^H = \p_n^- \Bigl ( \p_n^- (g\!-\! g^H )\, \delta n \Bigr )^H
\eeq
(this follows from the Hadamard variational formula).
Adding this to the result of Example 2, we get:
\beq\label{examp3}
\delta \oint_{\Gamma} f(z)\p_n^+ g_H (z) ds
= \oint_{\Gamma} \delta n (\nabla f , \nabla g )\, ds
- \oint_{\Gamma} \delta n \, \p_n^+ (f\! -\! f_H )\p_n^+ (g\! -\! g_H )\, ds,
\eeq
\beq\label{examp4}
\delta \oint_{\Gamma} f(z)\p_n^- g^H (z) ds
= \oint_{\Gamma} \delta n (\nabla f , \nabla g )\, ds
- \oint_{\Gamma} \delta n \, \p_n^- (f\! -\! f^H )\p_n^- (g\! -\! g^H )\, ds.
\eeq
As a simple corollary, we obtain:
\beq\label{examp5}
\delta \oint_{\Gamma} f\hat {\cal N} g  ds
= \oint_{\Gamma} \delta n \Bigl (\p_n^- (f\! -\! f^H )\p_n^- (g\! -\! g^H )-
\p_n^+ (f\! -\! f_H )\p_n^+ (g\! -\! g_H ) \Bigr ) ds.
\eeq

\paragraph{Example 4.} Here we show how to find variation of the external
conformal radius $r$ of the domain $\DD$. The external conformal radius is defined as
\beq\label{r1}
r=(w'(\infty )^{-1},
\eeq
where $w(z)$ is the conformal map from $\CC \setminus \DD$ onto the exterior of the unit circle
such that $w(\infty )=\infty$ and $w'(\infty )$ is real positive. The integral representation
of $\log r$ is
\beq\label{r2}
\log r =-\frac{1}{2\pi}\oint_{\Gamma}\log |w'(z)|\, |w'(z)| \, |dz|.
\eeq
A direct calculation using the Hadamard variational formula yields
\beq\label{r4}
\delta \log r = \frac{1}{2\pi}\oint_{\Gamma}|w'(z)|^2 \delta n(z) |dz|.
\eeq

\section*{Appendix F: Functional determinants}
\addcontentsline{toc}{section}{\hspace{6mm}Appendix F: Functional determinants}
\def\theequation{F\arabic{equation}}
\def\theHequation{\theequation}
\setcounter{equation}{0}

Given an elliptic operator $A$, one can define the $\zeta$-function $\zeta_A(s)$ as
\beq\label{zeta1}
\zeta_A(s)=\mbox{tr}'(A^{-s})={\sum\limits_{i}}'\lambda_i^{-s},
\eeq
where prime means summing over nonzero eigenvalues $\lambda_i$ only. The regularized 
determinant is defined as
\beq\label{zeta2}
{\det}'A = e^{-\zeta_A'(0)}.
\eeq
As a consequence, 
\beq\label{zeta3}
{\det}'(cA)=c^{\zeta_A(0)}{\det}'A
\eeq
for any constant $c$. 

Given a compact domain $\DD \subset \CC$, one can consider the Gaussian path integral
$$
\int [D\chi ]\exp \Bigl ( -\int_{\DD}|\nabla \chi |^2  d^2z\Bigr )=
\int [D\chi ]\exp \Bigl ( \int_{\DD}\chi \Delta \chi d^2z\Bigr )
$$
over a scalar field $\chi$ in $\DD$ with Dirichlet boundary condition $\chi =0$ on $\Gamma =\p \DD$.
In the r.h.s. $\Delta$ is the Laplace 
operator in $\DD$ acting on the space of 
functions with Dirichlet boundary conditions. The result is expressed through the functional
determinant of the Laplace operator:
\beq\label{path}
\int [D\chi ]\exp \Bigl ( \int_{\DD}\chi \Delta \chi d^2z\Bigr )=
C \Bigl (\det (-\Delta )\Bigr )^{-1/2},
\eeq
where $C$ is a divergent constant. The path integral (\ref{path}) implies a cut-off
at a small distance $\epsilon$. The result diverges as $\epsilon \to 0$. A finite part of the 
answer, obtained by subtracting the divergent terms in the limit $\epsilon \to 0$, is given
by analytic continuation of the $\zeta$-function $\zeta_{\Delta}(s)=\mbox{tr}((-\Delta )^{-s})$
to the point $s=0$. The operator $-\Delta$ in $\DD$ is known to have a discrete
spectrum with positive eigenvalues, so the trace is well-defined. The $\zeta$-regularized
determinant is defined as
\beq\label{detreg}
\det (-\Delta )=\exp \Bigl (-\zeta '_{-\Delta}(0)\Bigr )
\eeq
(see (\ref{zeta2})).
More generally, the definition of the determinant extends to the case when the metric
in the plane is not flat and $\Delta$ is the Laplace-Beltrami operator in this metric.

The determinant (\ref{detreg}) was calculated in \cite{Alvarez} in terms of the conformal 
factor of the metric. Let $\DD_0$ be a fixed reference domain with a background
metric $g_0$. The domain $\DD$ can be viewed as $\DD_0$ with the metric 
$g=g_0 e^{2\sigma}$ which differs from $g_0$ by the conformal factor $e^{2\sigma}$. Then
the Polyakov-Alvarez formula reads
\beq\label{PA}
\begin{array}{l}
\displaystyle{
\log \det \Bigl (-\Delta_{\DD}\Bigr )-\log \det \Bigl (-\Delta_{\DD_0}\Bigr )=
-\frac{1}{12\pi}\int_{\DD_0}\! d^2w \sqrt{g_0}g_0^{ab}\p_a \sigma \p_b \sigma -
\frac{1}{12\pi}\int_{\DD_0}\! d^2w \sqrt{g_0}R_0\sigma}
\\ \\
\displaystyle{\phantom{aaaaaaaaaaaaaaaaaaaaaaaaaaaaaaaa}
-\frac{1}{6\pi}\oint_{\p \DD_0}ds \kappa_0 \sigma +
\frac{1}{4\pi}\oint_{\p \DD_0}ds \p_n \sigma .}
\end{array}
\eeq
Here $R_0$ is the curvature of the metric $g_0$ in $\DD_0$ and $\kappa_0$ is the 
curvature of its boundary $\p \DD_0$.
In the main text we use this formula in the specific 
case when $\DD_0$ is the unit disk ${\sf U}$
with complex coordinate $w$ in the flat
metric and $\sigma (w) =\log |z'(w)|:=\phi (w)$, where $z(w)$ is the conformal map from 
the unit disk to the domain $\DD$. In this case 
$\sigma (w)=\phi (w)$ is harmonic in $\DD_0$, so the integral over $\DD_0$ reduces
to a boundary integral (by virtue of the Green theorem) 
and the Polyakov-Alvarez formula considerably simplifies:
\beq\label{PA1}
\log \det \Bigl (-\Delta_{\DD}\Bigr )-\log \det \Bigl (-\Delta_{{\sf U}}\Bigr )=
-\frac{1}{12\pi}\oint_{|w|=1} (\phi \p_n \phi +2\phi )|dw|
\eeq
(cf. (\ref{q1})).

For Laplace operators in 
non-compact exterior domains the definition outlined above does not work since the 
spectrum is continuous. To define determinants of Laplace operators in exterior domains,
a different approach is required. It was developed in \cite{HZ99}, where the 
determinant $\det (-\Delta_{\cc \setminus \DD})$ of the Laplace operator
in $\CC\setminus \DD$ was defined in terms of scattering on 
the compact ``obstacle'' ${\sf D}$. Here we will not go into details and only mention
that the term ``determinant'' for the object introduced by this definition
is justified by the ``surgery formula''
\beq\label{E5}
\log \det (-\Delta_{\DD})+\log \det (-\Delta_{\cc \setminus \DD})+\log {\det} ' \hat {\cal N}
=\log P +\mbox{const}
\eeq
proved in \cite{HZ99}. Here $\hat {\cal N}$ is the Neumann jump operator 
on $\Gamma =\p \DD$ and $P$ is the 
perimeter of $\p \DD$. 

Let us give a simple heuristic explanation of the surgery formula using formal
Gaussian path integrals over scalar fields. We start with the path integral
$$
Z_{\cc}=\int [D\chi ]\exp \Bigl ( -\int_{\cc}|\nabla \chi |^2  d^2z\Bigr )
$$ 
which does not depend on the contour and write
$$
\int_{\cc}|\nabla \chi |^2 d^2z =\int_{\DD}|\nabla \chi |^2d^2z 
+\int_{\cc \setminus \DD}|\nabla \chi |^2 d^2z.
$$
Let $f(z)$, $z\in \Gamma =\p \DD$ be the function $\chi (z)$ restricted
to the contour $\Gamma$. We can represent the field $\chi$ as
$$
\chi (z)=\left \{ \begin{array}{l}
f_H(z)+\chi_{\rm in}(z), \quad z\in \DD ,
\\ \\
f^H(z)+\chi_{\rm out}(z), \quad z\in \CC \setminus \DD ,
\end{array} \right.
$$
where $\chi_{\rm in}$, $\chi_{\rm out}$ are fields in the domains $\DD$ and 
$\CC \setminus \DD$ respectively with Dirichlet boundary conditions. 
A simple computation using the Green theorem shows that
$$
\int_{\cc}|\nabla \chi |^2 d^2z =\int_{\DD}|\nabla \chi_{\rm in} |^2 d^2z+
\int_{\cc \setminus \DD}|\nabla \chi_{\rm out} |^2 d^2z +\oint_{\Gamma}
f \hat {\cal N}f ds
$$
and, therefore, 
$$
Z_{\cc}=\int [D\chi_{\rm in}][D\chi_{\rm out}][Df]
\exp \left (-\int_{\DD}|\nabla \chi_{\rm in} |^2 d^2z
-\int_{\cc \setminus \DD}|\nabla \chi_{\rm out} |^2 d^2z-
\oint_{\Gamma}
f \hat {\cal N}f ds\right )
$$
$$
=C P^{1/2} (\det (-\Delta_{\DD}))^{-1/2} (\det (-\Delta_{\cc\setminus \DD}))^{-1/2} 
({\det }' \hat {\cal N})^{-1/2}
$$
with some contour-independent constant $C$. 
This gives a ``physical'' explanation of (\ref{E5}). The appearance of the factor $P^{1/2}$
is due to the zero mode of the operator $\hat {\cal N}$.

As is pointed out in \cite{HZ99}, the problem of defining the 
determinant of exterior Laplacian could be solved by placing the contour
on the sphere ${\sf S}^2$.
Let $g_0$ be 
be a metric in the plane of the form
$$
g_0(z)=\frac{g_{\rm flat}(z)}{f(|z|^2)}, \quad z\in \CC ,
$$
where $g_{\rm flat}$ is the standard flat metric and the function $f(x)$ is equal to 1
for $x\leq R$ and $f(x)=x^2$ for $x\geq R'$ ($R'>R$). In other words, $g_0$ is the same
as the standard metric on the disk ${\sf B}_{R,0}$ of radius $R$ centered at the origin
while its behavior at infinity means that $g_0$ extends to a smooth metric on ${\sf S}^2$
regarded as the compactification of $\CC$ by adding the point at infinity. This compactifies
$\CC \setminus \DD$ to a domain $\DD ^c \subset {\sf S}^2$. We assume that 
$0\in \DD$ and $\DD \subset {\sf B}_{R,0}$. 
We can compare (\ref{E5}) with the surgery formula from \cite{BFK92} applied for
${\sf S}^2$ with the metric $g_0$:
\beq\label{E6}
\log \det (-\Delta_{\DD})+\log \det (-\Delta_{\DD ^c, g_0})+\log {\det}' \hat {\cal N}=
\log {\det}' (-\Delta_{{\sf S}^2, g_0})+\log P +\mbox{const} .
\eeq
In particular, we can specialize it for the case when $\DD$ is the unit disk ${\sf U}$
with the boundary $\p {\sf U}={\sf S}^1$ (the unit circle):
\beq\label{E7}
\log \det (-\Delta_{\sf U})+\log \det (-\Delta_{{\sf U}^c, g_0})+
\log {\det}' \hat {\cal N}_{{\sf S}^1}=
\log {\det}' (-\Delta_{{\sf S}^2, g_0}) +\mbox{const} .
\eeq
Here ${\sf U}^c$ is the complement to the unit disk in the metric $g_0$. Note that
$\log \det (-\Delta_{\sf U})$ and $\log {\det}' \hat {\cal N}_{{\sf S}^1}$ are
universal constants. 
Now, summing (\ref{E5}) and (\ref{E7}) and subtracting (\ref{E6}), we obtain the 
relation
\beq\label{E8}
\log \det (-\Delta_{\cc \setminus \DD})=\log \det (-\Delta_{\DD ^c, g_0})-
\log \det (-\Delta_{{\sf U}^c, g_0}) +\mbox{const} .
\eeq
In order to find the quantity in the r.h.s., one can apply the Polyakov-Alvarez formula
(\ref{PA}) with $\DD _0 ={\sf U}^c$ with the background reference metric $g_0$
and $\sigma (w)=\log |z_{\rm ext}'(w)|:=\phi_{\rm ext}(w)$. Again, $\phi_{\rm ext}$ 
is harmonic in ${\sf U}^c$ and the first term in the r.h.s. of (\ref{PA}) reduces to
a contour integral. Tending $R\to \infty$, we see that the limit of the second term is $0$
and we obtain the result
\beq\label{E9}
\log \det \Bigl (-\Delta_{\cc \setminus \DD}\Bigr )=\frac{1}{12\pi}
\oint_{|w|=1} (\phi_{\rm ext}\p_n \phi_{\rm ext}+2\phi_{\rm ext})|dw| +\mbox{const}
\eeq
(cf. (\ref{q1a})). 

\section*{Acknowledgments}


We thank L. Friedlander, V. Kalvin, M. Putinar, and S. Zelditch for discussions and
pointing out some references. 
The work of A.Z. has been funded within the framework of the
HSE University Basic Research Program. The work of P.W. was supported by the NSF 
under the Grant NSF DMR-1949963.


\begin{thebibliography}{99}

\addcontentsline{toc}{section}{\hspace{6mm}References}

\bibitem{Dyson}
F. Dyson, {\it Statistical theory of the energy
levels of complex systems I, II}, J. Math. Phys. {\bf 3}
(1962) 140--156, 157--165.

\bibitem{Forrester}
P. Forrester, {\it Log-gases and random matrices}, London Math. Soc. Monographs, 
Princeton University Press, 2010.

\bibitem{Jancovici} 
B. Jancovici, {\it Classical Coulomb systems: 
Screening and correlations revisited},
Journal of Statistical Physics {\bf 80} (1995) 445-459.



\bibitem{WZ03} P. Wiegmann and A. Zabrodin,
{\it Large scale correlations in normal
non-Hermitian matrix ensembles}, J. Phys. A {\bf 36} (2003)
3411--3424.

\bibitem{WZ03a} P. Wiegmann and A. Zabrodin,
{\it Large $N$ expansion for normal and complex matrix ensembles}, 
in: Frontiers in Number Theory, Physics and Geometry I, P.Cartier, B. Julia, P. Moussa,
P. Vanhove, eds., 213--229, Springer, 2006.


\bibitem{WZ06} A. Zabrodin and P. Wiegmann,
{\it Large $N$ expansion for the 2D Dyson gas}, 
J. Phys. A {\bf 39} (2006) 8933--8964.

\bibitem{Z04} A. Zabrodin, {\it Matrix models and growth processes:
from viscous flows to the quantum Hall effect},
in: ``Applications of Random Matrices in
Physics'', pp. 261--318, Ed. E. Brezin et al, Springer, 2006, arXiv:hep-th/0412219.



\bibitem{Landafshitz}
L. Landau and E. Lifshitz, Course of Theoretical Physics, Volume 5,
{\it Statistical Physics, part 1}, 2nd revised and enlarged edition, Pergamon Press, 1969.

\bibitem{Polyakov}
A. Polyakov, {\it Quantum geometry of bosonic strings}, Phys. Lett. {\bf B103}
(1981) 207--210.

\bibitem{Alvarez}
O. Alvarez, {\it Theory of strings with boundaries: 
Fluctuations, topology and quantum geometry}, Nucl. Phys. {\bf B216} (1983)
125--184.

\bibitem{VDOP82}
B. Durhuus, P. Olesen and J.L. Petersen, {\it 
Polyakov's quantized string with boundary terms}, Nucl. Phys. {\bf B198}
(1982) 157--188.

\bibitem{ADFO86}
J. Ambjorn, B. Durhuus, J. Fr\"olich and P. Orland, {\it 
The appearance of critical dimensions in regulated string theories}, 
Nucl. Phys. {\bf B270} (1986)
457--482.

\bibitem{OPS88}
B. Osgood, R. Phillips and P. Sarnak, {\it Extremals of determinants of Laplacians},
J. Funct. Anal. {\bf 80} (1988) 148--211.

\bibitem{HZ99}
A. Hassel and S. Zelditch, {\it Determinants of Laplacians in exterior domains},
IMRN (1999) 971--1004, arXiv:math.AP/0002023.

\bibitem{BFK92}
D. Burghelea, L. Friedlander and T. Kappeler, {\it Mayer-Vietoris formula for determinants
of elliptic operators}, J. Funct. Anal. {\bf 107} (1992) 34--65.

\bibitem{Schiffer} M. Schiffer, {\it The Fredholm eigen values
of plane domains}, Pacific J. Math. {\bf 7} (1957) 1187--1225.

\bibitem{TT04}
L.-P. Teo and L. Takhtajan,
{Weil-Petersson metric on the universal Teichmuller space II. Kahler potential and period mapping}, arXiv:math/0406408.

\bibitem{EW91} J. Edwards and S. Wu, {\it Determinant of the Neumann operator on
smooth Jordan curves}, Proc. AMS {\bf 111} (1991) 357--363.

\bibitem{GG07} C. Guillarmou and L. Guilop\'e, {\it The determinants of the 
Dirichlet-to-Neumann map for surfaces with boundary}, IMRN, {\bf 2007} (2007) rnm099,
arXiv:math/0701727.

\bibitem{UT84} 
K. Ueno and K. Takasaki, {\it Toda lattice hierarchy}, Advanced Studies in Pure 
Mathematics {\bf 4} (1984) 1--95.


\bibitem{Taylor} M. Taylor, {\it Partial differential equations}, vol. II, {\it 
Qualitative studies of linear equations}, second edition, Springer, 1996.

\bibitem{MS17} Y. Miyanishi and T. Suzuki, {\it Eigenvalues and eigenfunctions of 
double layer potentials}, Trans. American Math. Soc. {\bf 369} (2017) 8037--8059.

\bibitem{AKMP20} K. Ando, H. Kang, Y. Miyanishi and M. Putinar, {\it Spectral 
analysis of Neumann-Poincar\'e operator}, arXiv:2003.14387.

\bibitem{KPS07} D. Khavinson, M. Putinar and H. Shapiro, {\it Poincar\'e's variational
problem in potential theory}, Arch. Rational Mech. Anal. {\bf 185} (2007) 143--184.

\bibitem{KV95} M. Kontsevich and S. Vishik, {\it Geometry of determinants of elliptic operators}, 
In: Gindikin S., Lepowsky J., Wilson R.L. (eds), Functional Analysis on the Eve of the 
21st Century, Progress in Mathematics, vol. 131/132 (1995).

\bibitem{Fried} L. Friedlander, {\it On multiplicative properties of determinants},
arXiv:1801.10606.

\bibitem{Johansson21}
K. Johansson, {\it Strong Szego theorem on a
Jordan curve}, arXiv:2110.11032.


\end{thebibliography}
\end{document}